\documentclass[12pt]{article}
\pdfoutput=1

\usepackage{graphics}
\usepackage{amsmath,amsfonts}

\usepackage{amsmath}
\usepackage{amssymb}
\usepackage{amsthm}
\usepackage{psfrag}
\usepackage{graphicx}
\usepackage{color}
\usepackage{mathtools}
\usepackage{hyperref}
\usepackage{array}
\usepackage{cite}

\newcommand{\beq}{\begin{equation}}
\newcommand{\eeq}{\end{equation}}
\newcommand{\beqa}{\begin{eqnarray}}
\newcommand{\eeqa}{\end{eqnarray}}
\newcommand{\bea}{\begin{eqnarray}}
\newcommand{\eea}{\end{eqnarray}}

\newcommand{\OO}{{\cal O}}

\DeclarePairedDelimiter\abs{\lvert}{\rvert}

\begin{document}

\begin{center}
\vspace{2.5cm}

\sc{\huge \bf Oscillons and Dark Matter}


\vspace{1cm}

Jan Oll\'e, Oriol Pujol\`as and Fabrizio Rompineve\\

\vspace{1cm}

{\it Institut de F\'{\i}sica d'Altes Energies (IFAE)\\ 
\normalsize\it The Barcelona Institute of  Science and Technology (BIST)\\
\normalsize\it Campus UAB, 08193 Bellaterra (Barcelona) Spain}\\

\end{center}


\begin{abstract}

\noindent Oscillons are bound states sustained by self-interactions that appear in rather generic scalar models. 
They can be extremely long-lived and 
have a built-in formation mechanism -- parametric resonance instability. 
These features suggest that oscillons can affect the standard picture of scalar ultra-light dark matter (ULDM) models. 
We explore this idea along two directions.
First, we investigate numerically oscillon lifetimes and their dependence on the shape of the potential. 
We find that scalar potentials that occur in well motivated axion-like models can lead to oscillons that live up to $10^8$ cycles or more. 
Second, we discuss the observational constraints on the ULDM models once the presence of oscillons is taken into account. 
For a wide range of axion masses, oscillons decay around or after matter-radiation equality and can thus act as early seeds for structure formation. We also discuss the possibility that oscillons survive up to today. In this case they can most easily play the role of dark matter. 

\end{abstract}

\newpage 

\tableofcontents

\newpage

\section{Introduction}

In our Universe, a variety of visible structures with different length scales form as aggregates of smaller constituents that are kept together by the gravitational force. This is the case for planets, stars, solar systems and galaxies.
The formation of bound objects is of course generic to any attractive interaction, and it is natural to wonder what would be the effect of self-interactions in the dark matter sector. 
Very much like gravity for visible matter, attractive self-interactions must lead to the formation of dark structures which can have important observational consequences.

Nontrivial bound states arise in the simplest field theory that one can imagine: a real scalar field with an attractive self-interaction. The bound objects in this case are then usually called  {\it oscillons}. 
Interestingly, for very light scalar mass $m$, this class of models also offers a viable and well motivated dark matter candidate, what is usually dubbed scalar Ultra Light Dark Matter (ULDM)~\cite{Turner:1983he, Press:1989id, Sin:1992bg, Hu:2000ke, Goodman:2000tg, Peebles:2000yy, Amendola:2005ad, Schive:2014dra, Hui:2016ltb}. In this framework, then, dark matter can form by itself structures with rather definite properties (mass, size, lifetime).
This paper is devoted to the study of how self-interactions affect the usual picture of ULDM, where they are mostly neglected even though in many models they are present or implicit.
The most important effect of the attractive self-interactions is that at some point  \emph{oscillons} arise as part of the dark sector.

\begin{figure}[t]
\centering
\includegraphics[width=0.7\textwidth]{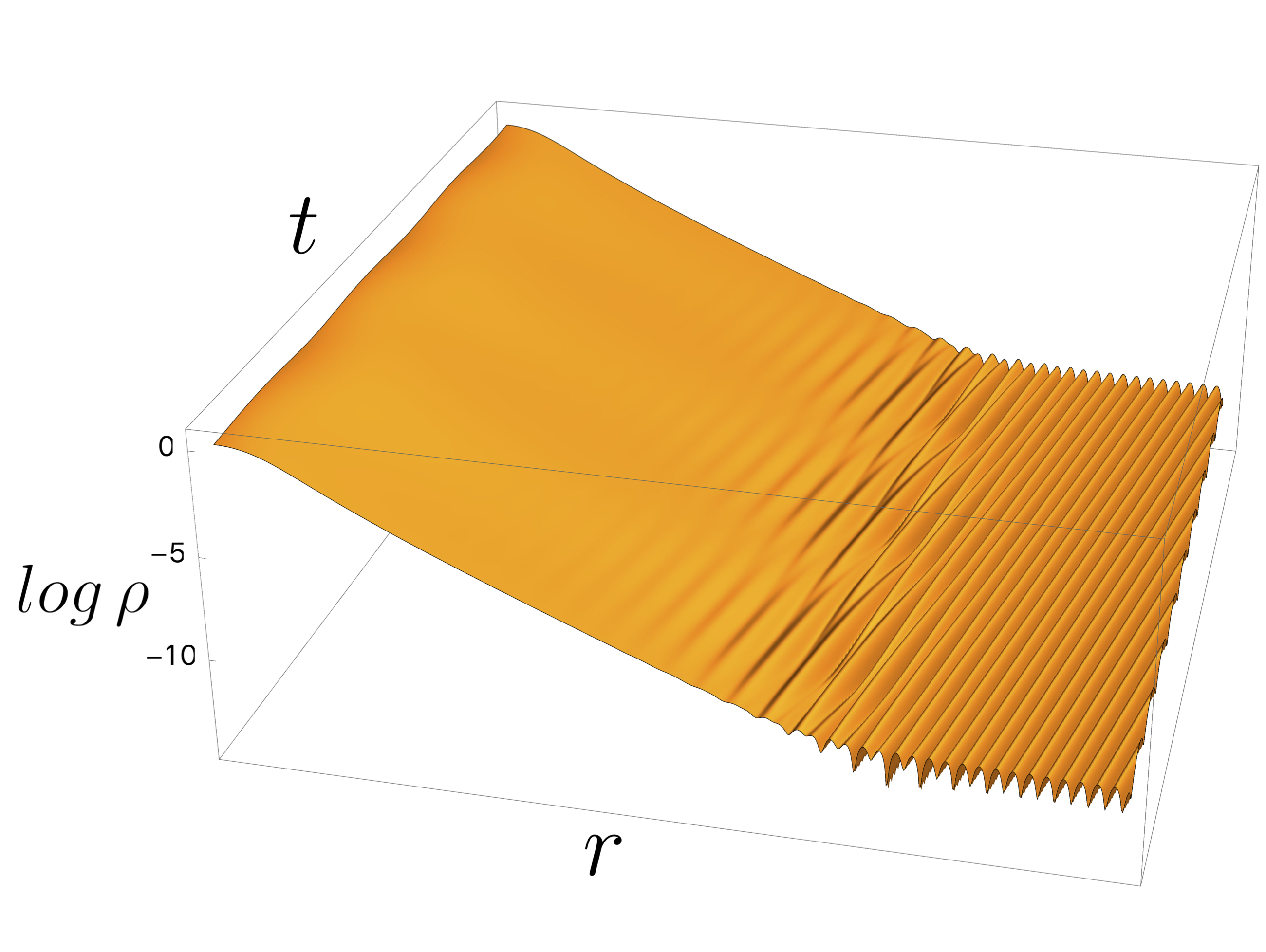}
\caption{\it Energy density profile of an oscillon: away from the central core the profile decays exponentially, until radiated scalar waves become relevant. Inside the core, the energy density oscillates in time.}
\label{fig:oscillon}
\end{figure}

Oscillons are localized and oscillating configurations of a real scalar field, which are sustained solely by self-interactions~\cite{Bogolyubsky:1976yu, Gleiser:1993pt, Copeland:1995fq}. The most relevant ones are spherically symmetric $\phi=\phi(t,r)$, so one can present them as solutions of the Klein-Gordon equation in flat spacetime with a nonlinear potential,
\begin{equation}
\ddot{\phi}(t, r) - \partial^{2}_{r}\phi(t, r) -\frac{2}{r}\partial_{r}\phi(t, r) + V'(\phi(t,r))=0~. \label{eq:eom}
\end{equation}
Since a real scalar field carries no conserved charge, oscillons are unstable\footnote{A charged scalar field can also lead to localized non-topological objects, known as Q-balls~\cite{Coleman:1985ki, Kusenko:1997si}, which can be stable provided self interactions are attractive.}:  they lose energy even classically by emitting radiation in the form of scalar waves, and eventually decay into a `wave-packet' that diffuses away (we show the energy density profile of an oscillon with its radiated waves in Fig.~\ref{fig:oscillon}). Nevertheless, depending on the strength and type of self-interactions, the radiation rate can be extremely slow, thereby leading to large oscillon lifetimes $\tau\gg m^{-1}$ and making them potentially very interesting for cosmology.

There is a considerable amount of literature on the topic of oscillons treating the conditions for them to appear, how can they be understood as attractors of the dynamics and trying to understand what determines their longevity, see e.g.~\cite{Honda:2001xg, Kasuya:2002zs, Fodor:2006zs, Saffin:2006yk, Hindmarsh:2007jb, Fodor:2008es, Gleiser:2008ty, Fodor:2009kf, Amin:2010jq,  Hertzberg:2010yz, Amin:2011hj, Salmi:2012ta, Andersen:2012wg, Saffin:2014yka, Mukaida:2016hwd, Ibe:2019vyo, Gleiser:2019rvw, Dymnikova:2000dy} for a partial list of relevant work. 
For the purpose of elucidating their impact on cosmology, there are two basic requirements that guarantee that at some point oscillons will be present. On the one hand, the field needs to start out from a sufficiently inhomogeneous configuration. On the other hand, the scalar potential $V(\phi)$ needs to contain attractive self-interactions; more precisely, it should exhibit regions which are flatter than quadratic, i.e. $V'<m^{2}\phi$, otherwise self-interactions cannot sustain the localized overdensity. Both conditions are satisfied in a range of well-motivated models. The first example is the QCD axion model with $V(\phi)=m^2 F^2(1- \cos{(\phi/F)})$, which is known to exhibit oscillon solutions, also referred to as \emph{axitons}~\cite{Kolb:1993hw, Kolb:1994fi} (see also~\cite{Hogan:1988mp} for the formation of so-called \emph{miniclusters}, \cite{Vaquero:2018tib, Buschmann:2019icd} for recent simulations and \cite{Visinelli:2017ooc} for \emph{axion stars}). In this model, however, oscillons are not very long-lived: their lifetimes are typically `only' of order $\tau \sim 10^3~m^{-1}$. 

\begin{figure}[t]
\centering
\includegraphics[width=0.8\textwidth]{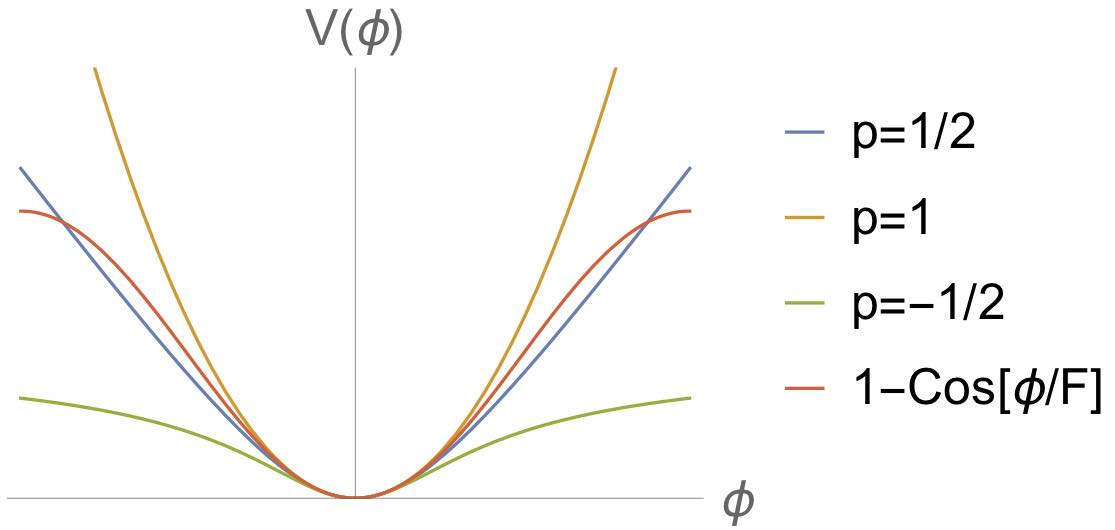}
\caption{\it Scalar potentials of the form~\eqref{eq:potmon}, compared to the standard quadratic ($p=1$) and cosine potentials for axions.}
\label{fig:potentials}
\end{figure}

Still, in the zoo of models that give rise to ultra-light scalars,  a much richer variety of potentials has been considered. For instance, in  so-called axion monodromy~\cite{Silverstein:2008sg, McAllister:2008hb} constructions arising from string theory, the scalar potential can asymptote to a power-law behaviour at large field values $V(\phi)\sim \phi^{2p}$, with a variety of exponents $p$ available in the market~\cite{Dong:2010in}.
Having in mind this class of axion models, we will consider as a benchmark a family of scalar potentials that allows to make contact with these UV models while being simple enough to do a careful numerical study of the oscillon longevity. Specifically, we will take
\begin{equation}
\label{eq:potmon}
V(\phi)=\frac{m^2 F^{2}}{2p}\left[\left(1+\frac{\phi^{2}}{F^{2}}\right)^{p}-1 \right]~,
\end{equation}
which is expected to lead to oscillons for $p<1$. In this parameterization, there are two rather distinct cases. 
When $0\leq p<1$, the potentials are of axion-monodromy form since for  $\phi\gg F$ the potential grows as a power law. For $p<0$ instead, the potential saturates at large field values and exhibits a plateau region, see Fig.~\ref{fig:potentials}.

The requirement to have initially inhomogeneous configurations is also automatically realized in this kind of ULDM model. The mechanism goes as follows: since the scalar is very light, it remains stuck because of Hubble friction during a large portion of the cosmological history. Assuming that the field is already present but subdominant during inflation, then the latter stretches it out to a very homogeneous value which does not need to coincide with the minimum of the potential. Well into the radiation era, the field is then going to roll down its potential and produce a homogeneous but time dependent background $\phi(t)$. Crucially, the homogeneous oscillations following from a potential of the form \eqref{eq:potmon} with $p<1$
lead to a parametric resonance instability of the vacuum fluctuations around $\phi(t)$: certain modes grow exponentially until the field has dumped a large fraction of its energy density into inhomogeneities. The condition that ensures that  oscillon solutions form also guarantees that the homogeneous configuration is unstable. Therefore potentials like \eqref{eq:potmon} (with $p<1$) are equipped with a {\em fragmentation mechanism} whereby the homogeneous configuration is transformed into an admixture of oscillons and scalar radiation. Numerical simulations of this process show that in many cases the initial energy density goes in equal proportions to oscillons and small amplitude scalar waves~\cite{Lozanov:2017hjm, Kitajima:2018zco}.

These appealing features of ULDM models with potentials \eqref{eq:potmon} motivate the present study, which proceeds along two different directions. First, we focus on the formation and evolution of oscillons. To this aim, we perform a linear Floquet analysis of the potentials~\eqref{eq:potmon} and describe under which conditions parametric resonance provides the right ground to form oscillons. We then numerically follow the evolution of a single oscillon configuration, according to \eqref{eq:eom}. By also adding Hubble friction, we show that oscillons are essentially unaffected by the cosmological expansion, as they form with sizes which are significantly smaller than the Hubble radius at the time of formation. Thus, they can be considered as bound objects which are decoupled from the Hubble flow, very much like any gravitationally bound system in the late Universe.
We aim in particular at establishing oscillon lifetimes and their dependence on the exponent $p$ in~\eqref{eq:potmon}. Very interestingly, while we succeed in doing so for $p\geq 0$, we are only able to put a lower bound on the lifetime for $p<0$, due to our current numerical reach.
The oscillons which we investigate are the longest-lived ever encountered, with lifetimes of order $10^{8}~m^{-1}$ and even larger in the case $p<0$.

These findings motivate the second part of our study, which deals with the observational impact of oscillons for dark matter models. Indeed, while oscillons form during the radiation dominated era from scalar field oscillations, for sufficiently small scalar masses they live up to and beyond the epoch of matter-radiation (MR) equality. Thus, they can potentially leave an important imprint on cosmology, either by serving as seeds for early structures or even by possibly making up a significant fraction of the dark matter today. These possibilities are especially likely to occur in the mass range which has been recently investigated in the context of fuzzy DM models~\cite{Hu:2000ke, Hui:2016ltb}. Here, we try to make contact with another peculiarity of this class of DM candidates, by proposing oscillons as early seeds for the well-known solitons of ULDM, which are however sustained by gravity~(see~\cite{Schiappacasse:2017ham, Hertzberg:2018lmt, Deng:2018jjz, Bar:2018acw} for more recent studies), in contrast with the objects which we focus on in this work. We thus describe constraints on different scenarios with dark matter made of two components, at least during a part of the cosmological history: field oscillations around the minimum of the potential and MACHO-like oscillons, with masses which increase as $F$ increases and $m$ decreases and span the range $10^{-12}~M_{\odot}$ to $10^{8} M_{\odot}$. The energy of these oscillons is concentrated in a region of size set by the inverse scalar mass, ranging between asteroid and galactic core scales as $m$ decreases. One of the main points of this paper is thus that future numerical and analytic studies of scalar DM, and in particular of ULDM (see e.g. the simulations of~\cite{Schive:2014dra}), should take into account the presence of overdensities on the scales of the oscillon size. This may lead to important changes in current constraints on these scenarios (see e.g.~\cite{Marsh:2015xka}).

Before moving on to the main content, let us put our work into context. The potentials~\eqref{eq:potmon} with $p>0$ have been studied in the framework of inflation and arise in string-theoretical constructions of axion inflation~\cite{Silverstein:2008sg, McAllister:2008hb}. In some cases these potentials can be seen as representing theories with multi-branched potentials (see~\cite{Witten:1980sp, Witten:1998uka} and~\cite{Dvali:2005an, Kaloper:2008fb, Kaloper:2011jz, Yonekura:2014oja}). Parametric instability and formation of oscillons during reheating after inflation has also been investigated in these setups~\cite{Amin:2010jq, Amin:2011hj, Antusch:2017flz, Lozanov:2017hjm} (see also~\cite{Zhou:2013tsa, Amin:2018xfe} for the emission of gravitational waves from oscillons). Potentials with a plateau at large field values can also be obtained and are also used for inflation~\cite{Kallosh:2013hoa, Kallosh:2013tua, Kallosh:2013yoa, Nomura:2017ehb, Nomura:2017zqj, Landete:2017amp}. Their instabilities and the generation of oscillons have been studied in~\cite{Hong:2017ooe, Kitajima:2018zco, Lozanov:2019ylm} and in~\cite{Fukunaga:2019unq}, which also discuss some aspects of plateau-like potentials for ULDM models. 
Furthermore, in the interesting proposal \cite{Cotner:2018vug}, oscillons play a key role for dark matter, as `mothers' to primordial black holes: at end of inflation oscillons from the inflaton field may dominate for long enough to form black holes which in the end account for dark matter.

This paper is organized as follows: in Sec.~\ref{sec:oscillonbio} we study the conditions under which parametric resonance is efficient~(\ref{sub:birth}); we numerically determine oscillon lifetimes~(\ref{sec:lifetime}); finally we discuss the decay of an oscillon and its possible transition to a gravitationally-sustained soliton~(\ref{sub:legacy}). We devote Sec.~\ref{sec:pheno} to the observational implications of long-lived oscillons, separately discussing the cases of decay after~(\ref{sub:after}) and before~(\ref{sub:before}) MR equality. We discuss the possibility for oscillons to survive until today in Sec.~\ref{sec:perpetual_oscillons}. Finally, we offer our conclusions in Sec.~\ref{sec:conclusions}.

\section{Life and death of an oscillon}
\label{sec:oscillonbio}

\subsection{Birth}
\label{sub:birth}

Oscillons are born out of inhomogeneous field configurations, which can originate from different mechanisms. On the one hand, one can consider a (pseudo)scalar field which is inhomogeneous from the very beginning of its cosmological history, as a result e.g.~of a phase transition which leaves behind topological defects. This is the case of the QCD axion if the PQ symmetry is broken after inflation (see~\cite{Kolb:1993hw, Kolb:1994fi}).

Alternatively, the scalar field can originally be homogeneous in the reheated universe and fragment into oscillons later during its evolution. This possibility is particularly relevant for axion fields with very large $F$, since in this case the PQ symmetry is broken during inflation, and will be the focus of this section. Starting from field values of $O(F)$, the time evolution of the homogeneous field then follows the nonlinear equation 
\begin{equation}
\label{eq:background}
\ddot\phi(t)+3\,H\,\dot\phi(t)+V'(\phi(t))=0.
\end{equation}
The way how the homogeneous field oscillations settle-in depends on the form of $V(\phi)$ (see e.g.~\cite{Masso:2005zg} and references therein). 
Let us denote by $\phi_{0}$ the amplitude of oscillations once they start. 
A good criterium for the Hubble rate at the beginning of the field oscillations is given by~\cite{Kitajima:2018zco}
\begin{equation}
\label{eq:hosc}
H_{\text{osc}}\simeq \sqrt{\left\lvert\frac{V'(\phi_{0})}{\phi_{0}}\right\rvert}.
\end{equation}
The latter equation is a natural replacement for the criterium $H_{\text{osc}} \simeq m$ for general nonlinear potentials. Notice that for $\phi_{0}\gg F$ or for potentials which are very flat beyond $\phi\gtrsim F$, one has $H_{\text{osc}}\ll m$. Otherwise, $H_{\text{osc}}\lesssim m$, as usual.

Crucially for us, potentials of the form~\eqref{eq:potmon} support the phenomenon of parametric resonance: fluctuations of the initially homogeneous field can grow exponentially as the background field undergoes slow oscillations. Once a given mode becomes large enough, its further growth is prevented by backreaction of other modes and the resonant phase stops, leaving behind the desired highly inhomogeneous field configuration.

Here we want to assess the conditions under which parametric resonance can be strong enough for the potentials \eqref{eq:potmon} (see also~\cite{Lozanov:2017hjm} for a recent study in the inflationary context). Let us then perform a linear analysis of the fluctuations, defined by $\phi\equiv \phi(t)+\delta\phi(\mathbf{x}, t)$.
The equations of motion of the linear fluctuations in Fourier space are 
\begin{equation}
\label{eq:fluctuations}
\delta\ddot{\phi}_{k}(t)+3H\delta\phi_{k}(t)+\left(\frac{k^{2}}{a^{2}}+V''(\phi(t))\right)\delta\phi_{k}(t)=0,
\end{equation}
where $k=\abs{\mathbf{k}}$ and $a\equiv a(t)\sim t^{1/2}$. Let us first neglect Hubble friction, setting $H=0$ and $a=1$ in \eqref{eq:background} and \eqref{eq:fluctuations}. The motion of the background field in the potential \eqref{eq:potmon} is periodic with period $T$ for any value of $p$ and in particular harmonic for $p=1$. The second derivative of the potential, which acts as source in the equation for the fluctuations, is then also periodic, with period $T/2$. Therefore, \eqref{eq:fluctuations} is in the form of \emph{Hill's equation} (see e.g.~\cite{whittaker1996course} and the appendix of~\cite{Amin:2011hu} for a review in the context of reheating). The solution to this equation is of the form $\delta\phi_{k}\sim e^{\mu_{k} t}P(t)$, where $P$ is periodic with period $T/2$ and $\mu_{k}$ is in general a complex number. If $\Re(\mu_{k})>0$, then the $k$-th mode grows exponentially, signaling an instability. This occurs if $V(\phi)$ is flatter than quadratic in some region of field space. Once $\delta\phi_{k}\sim \phi$, the linear analysis breaks down and one needs to include backreaction of the other modes.

\begin{figure}[t]
\centering
\includegraphics[width=\textwidth]{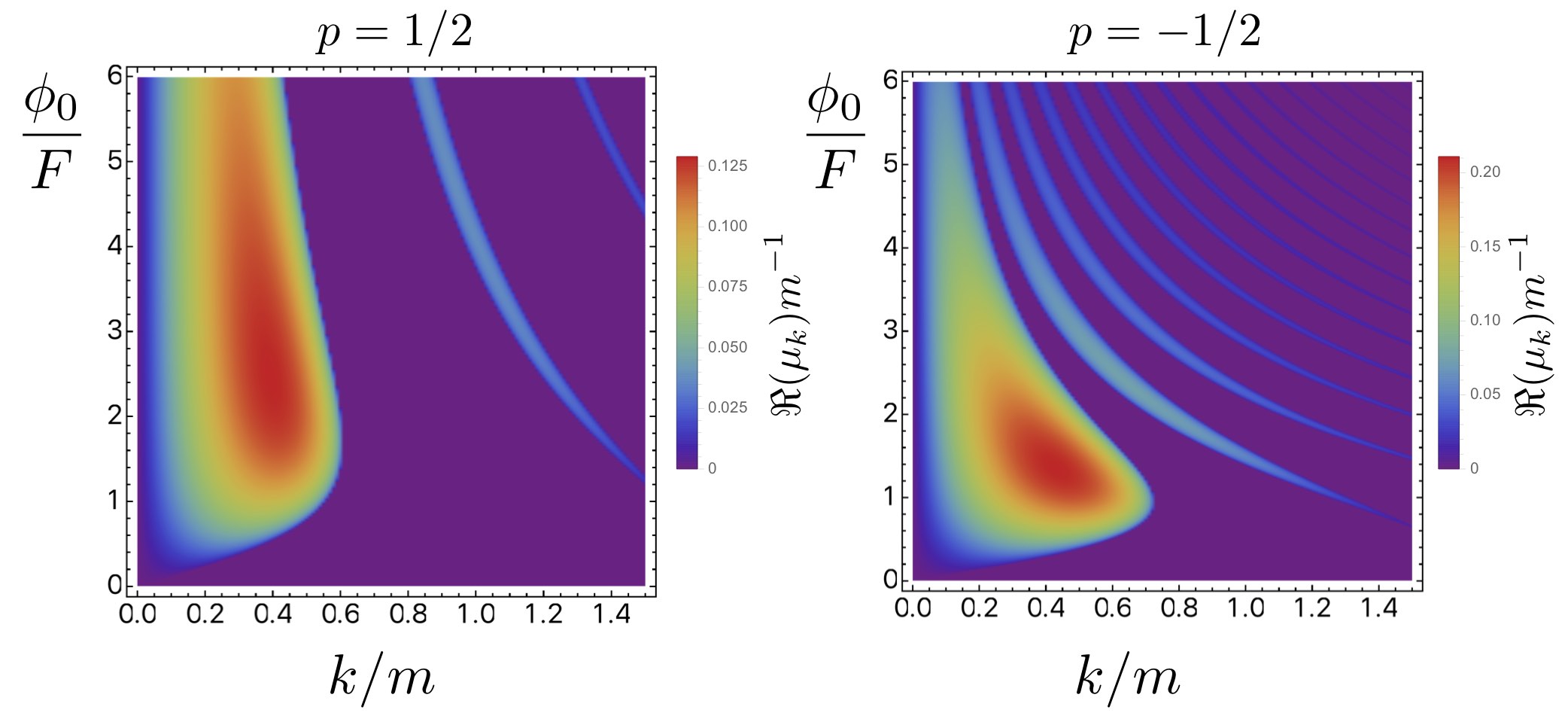}
\caption{\it Floquet diagrams for the monodromy potential \eqref{eq:potmon} with $p=1/2$ (left) and $p=-1/2$ (right). These plots are obtained by setting $H=0, a=1$ in \eqref{eq:background} and \eqref{eq:fluctuations}.}
\label{fig:floquet05}
\end{figure}

For our purposes, the strength of the resonance can be understood by looking at the values of $\mu_{k}$ as a function of $k$ and $\phi_{0}$ (see the appendix of~\cite{Amin:2011hu} for a computational method), plotted in Fig.~\ref{fig:floquet05} for $p=\pm 1/2$. In both cases the largest growth occurs in a broad band around $k\simeq m/2$ and $\phi_{0}\sim 2 F$. The case $p=-1/2$ is characterized by many more bands than $p=1/2$ and the region of parameter space where $\mu_{k}$ is largest is smaller compared to $p=1/2$. In both cases, the maximal size of the Floquet exponent is $O(0.1)m$.

Let us now restore Hubble friction in \eqref{eq:background} and \eqref{eq:fluctuations}. This leads to two effects: firstly, it introduces the time scale $H^{-1}$ to which we have to compare $\mu_{k}^{-1}$. If $\mu_{k}^{-1}/H^{-1}\ll 1$ then strong resonance can indeed occur. Otherwise, the field remains approximately homogeneous. Secondly, each mode $k$ now redshifts as $a^{-1}$, thereby moving across the parameter space in Fig.~\ref{fig:floquet05}. If a given $k$ spends enough time in the bands of resonant amplification, then the field will be inhomogenized. 

Let us first focus on the comparison between $\mu_{k}^{-1}$ and $H^{-1}$. At the time $t_{\text{osc}}$ defined by \eqref{eq:hosc}, the ratio $\mu_{k}(\phi_{0})/H_{\text{osc}}$ may or may not be sufficiently large to induce strong resonance.  However, as the Universe expands the Hubble rate decreases as $t^{-1}$, while the background amplitude decreases more slowly, in particular as $\sim a^{-\frac{3}{2}(1+w)}\sim t^{-3/4(1+w)}$ with $w\lesssim 0$. Therefore, the ratio $\mu_{k}(\phi(t))/H(t)$ can become large and parametric resonance efficient at some time $t_{\text{res}}>t_{\text{osc}}$. Numerically, we find that for $p=\pm 1/2$, $\mu_{k}(\phi)/H\gtrsim 10$ in the broad bands of Fig.~\ref{fig:floquet05} for $\phi_{0}/F \gtrsim 10$, signaling that strong resonance does occur for mildly large initial values of the axion field.\footnote{This means that the field starts its oscillations at $\phi_{0}$, then redshifts due to Hubble friction and crosses the resonance bands shown in Fig.~\ref{fig:floquet05}.} This is in contrast with the case of parametric resonance during reheating, where one needs $\phi_{0}\gtrsim100 F$ to have efficient resonance \cite{Amin:2011hj, Amin:2019ums}.

\begin{figure}[t]
\centering
\includegraphics[width=\textwidth]{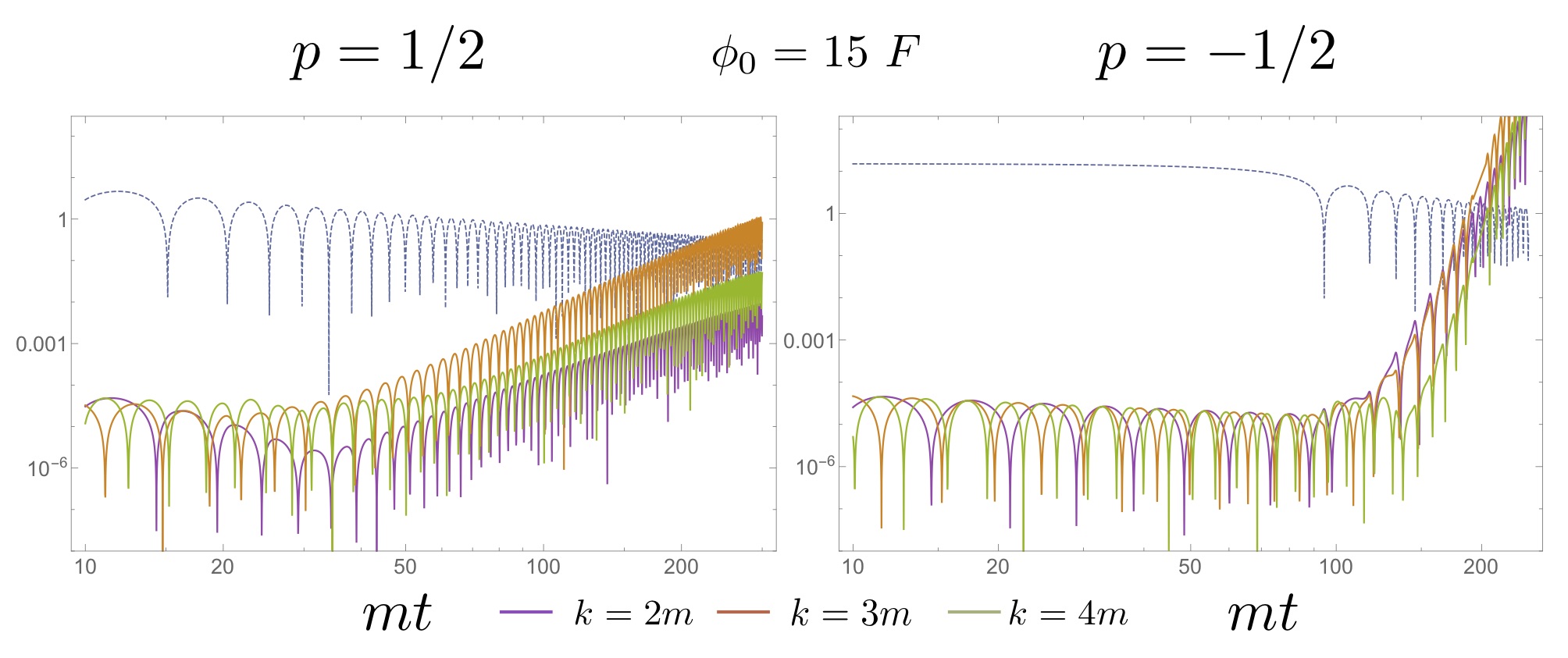}
\caption{\it LogLog plot of the absolute value of fluctuations $\delta\phi_{k}(t)/F$, with $k=2,3,4$ (thick curves) as a function of $t$. The dashed curve represents the evolution of the homogeneous background. Here we have fixed its initial value $\phi_{0}=15~F$ and $\delta\phi_{k}(t_{\text{osc}})=10^{-5}\phi_{0}$.}
\label{fig:modes15}
\end{figure}

Let us now address the second effect. To this aim, we compute numerically the evolution of some modes according to \eqref{eq:background} and \eqref{eq:fluctuations}. Because of the redshift, we now find that the most amplified modes have $k\gtrsim 2m$ and that $\phi_{\text{0}}\gtrsim 10$ is indeed enough for these modes to reach $\delta\phi_{k}\sim \phi$.
In Fig.~\ref{fig:modes15} we show the resonant amplification of three modes with initial amplitude $\delta\phi_{k}(t_{\text{osc}})=10^{-5}\phi_{0}$ for $\phi_{0}=15~F$ and $p=\pm 1/2$. We observe that strong resonance does indeed occur on a timescale $t_{\text{res}} \sim 100~m^{-1}$. At $mt\sim 200$, multiple modes have reached the same amplitude as the background and the linear analysis ceases to be reliable. The initial condition has been chosen in such a way that fluctuations are amplified to values which are close to $F$.

\begin{figure}[t]
\centering
\includegraphics[width=\textwidth]{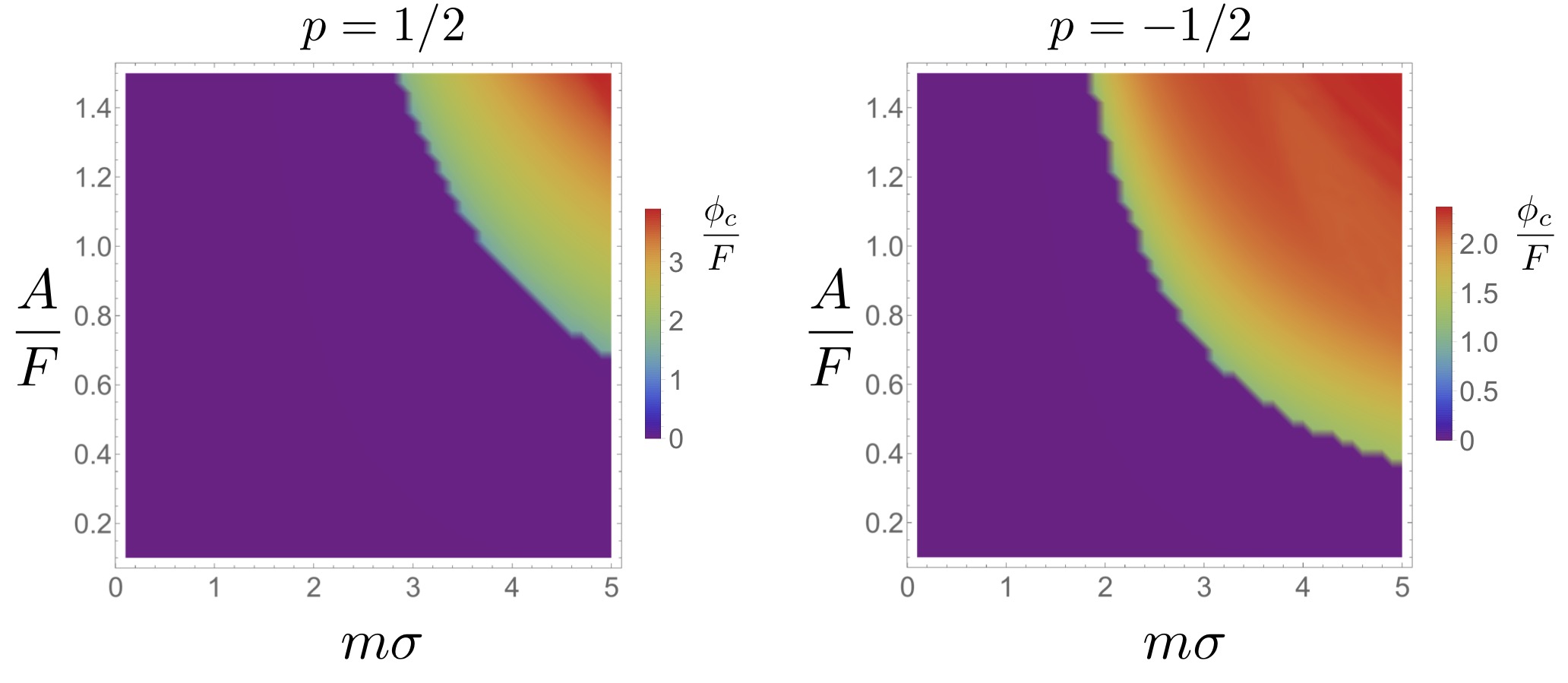}
\caption{\it Core amplitude at $t=10^{4}~m^{-1}$ of the localized field configuration initialized with \eqref{eq:profile}, as a function of $A$ and $\sigma$. The rainbow colored regions correspond to oscillon formation. In the purple region, axion lumps decay away soon after formation.}
\label{fig:cosh}
\end{figure}

This simple linear analysis probably overestimates the final values of the field fluctuations. However, the large values close to $F$ which we find are in principle not necessary to form oscillons, as we now explain. Parametric resonance generically leaves behind localized, approximately spherically symmetric overdensities, whose typical length scale is given by $2\pi/(k/a)$, with $k/a \lesssim m$ at the time of amplification, as can be seen from Fig.~\ref{fig:floquet05}. A convenient parametrization of the initial spatial profiles of such lumps, which also turns out to match well the final oscillon shapes, is given by
\begin{equation}
\label{eq:profile}
\phi(t=t_{\text{res}}, r)=\frac{A}{\cosh(r/\sigma)},
\end{equation}
where $\sigma\sim (k/a)^{-1}$ and $A\sim\delta\phi_{k}$ at the end of parametric resonance. In order to understand what ranges of $A$ and $\sigma$ lead to long lived configurations, i.e. oscillons, we have evolved the initial condition  \eqref{eq:profile} (with $\dot\phi(t_{res},r)=0$) according to the full non-linear equation of motion \eqref{eq:eom}, scanning over the profile initial amplitude and width. The results are shown in Fig.~\ref{fig:cosh}, where we plot the field amplitude $\phi_{c}$ at the lump core at the time $10^{4}~m^{-1}$ as we vary $A$ and $\sigma$. The crucial point is that localized configurations with sufficiently large initial amplitude and width evolve to  new configurations with larger amplitude (and smaller width), which is nothing but the early stage of an oscillon.
The exercise shown in Fig.~\ref{fig:cosh} is a way to track how the parametric resonance instability continues developing during the nonlinear regime for localized overdensities. Strikingly, even with very modest initial amplitudes $A \sim 10^{-1}  F$ the tendency to raise the core amplitude and form oscillons is clear (see~\cite{Gleiser:2019rvw} for a recent study along similar lines). This occurs for sufficiently large values of $\sigma$, such as those that we expect from parametric resonance.
In the next subsection, we will see that the new overdensity is stable across a time scale which is much larger than the one investigated in Fig.~\ref{fig:cosh}.\footnote{Fig.~\ref{fig:cosh} also makes manifest that the parametric resonance instability and the extreme longevity of oscillons are deeply interconnected: as the oscillon tries to decay it goes to a thicker configuration, which suffers more severely from the instability that tries to make it contract.}
Therefore, $\phi_{0} \sim 10 F$ is enough to generate long lived localized field configurations.

\subsection{Lifetime}
\label{sec:lifetime}

We now focus on the dynamics of a single oscillon, which is characterized by the very slow classical radiation of scalar waves. This can be investigated by numerically evolving the oscillon profile according to the full non-linear equation of motion \eqref{eq:eom}.
In general oscillons also emit quantum radiation, that is, they can decay by quantum mechanical processes. However, quantum radiation is always suppressed by coupling constants. Assuming that the couplings to any other particles is small, then, one can consistently neglect this radiation. Quantum emission of scalar quanta is also suppressed by $m^2/F^2$ \cite{Hertzberg:2010yz, Dvali:2017ruz}, which is very small in the ULDM setup which we focus on. This justifies our focus on classical processes.

Oscillons can be understood as being attractors of inhomogeneous field configurations \cite{Gleiser:2009ys}. This can already be appreciated in Fig.~\ref{fig:cosh}, where by starting with a profile according to \eqref{eq:profile}, one still finds a localized solution after time $t = 10^4 \, m^{-1}$. Of special relevance is the fact that the field amplitude at the core $\phi_c$ after such time is in general $\phi_c > A$, meaning the field has reorganized its energy density until it has found another profile which is preferred by the dynamics. Such a profile is an attractor of the dynamics and is none other than the oscillon. By means of a simple fit, one finds that the shape of this object is very well described by \eqref{eq:profile}, albeit with different $A$ and $\sigma$ than the ones it started with. It is also worth mentioning that the attractor profile is insensitive to specific initial conditions as long as the field is sufficiently inhomogeneous. This result motivates starting our numerical evolution with the profile~\eqref{eq:profile}.

For random $A$ and $\sigma$, however, one would have to spend some computational time until the field relaxes into the real oscillon configuration. In order to reduce this waiting time we use a shooting algorithm that allows us to find the $\sigma$ closest to an oscillon configuration once an $A$ has been fixed. We are thus left only with the choice of the initial amplitude $A$, which could result in different lifetimes. 

\begin{figure}[t]
\centering
\includegraphics[width=0.8\textwidth]{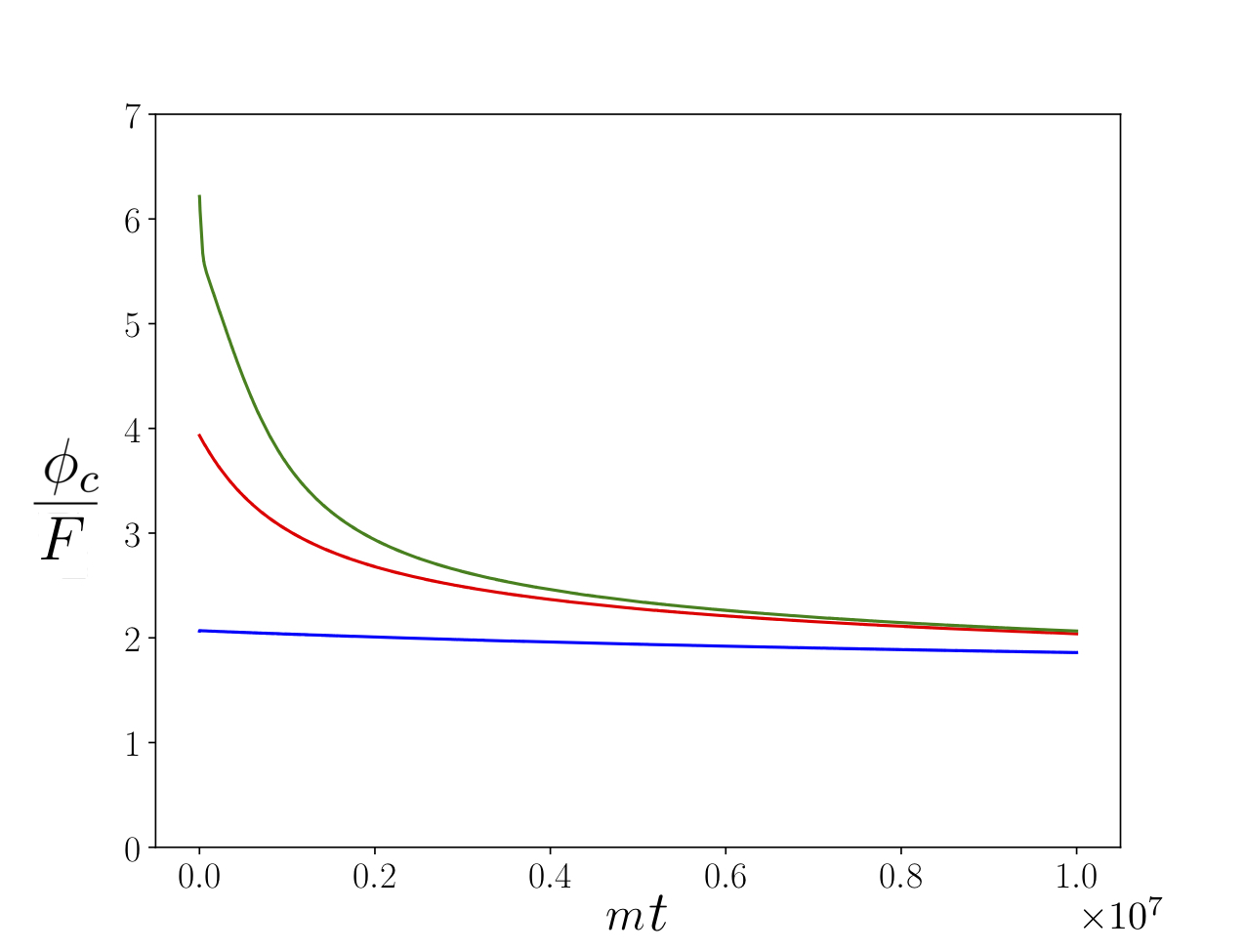}
\caption{\it Time evolution of the mean of the absolute value of the field at the center of the oscillon, $\phi_c$, for $p = 1/2$ and three different initial conditions.}
\label{fig:birth_attractor}
\end{figure}

Intuitively, the smaller $A$, the smaller the lifetime. However, starting with a larger amplitude does not result in a dramatic longevity increase. This is illustrated in Fig.~\ref{fig:birth_attractor} for the case $p = 1/2$ and manifestly illustrates the existence of a basin of attraction to which inhomoegenous field configurations quickly tend to. Although we have used $p = 1/2$ as a representative example, we have checked that this result is model-independent.

Together with the core field value $\phi_c$, the defining properties of an oscillon are its radius $R_{\text{osc}}$; its total mass $M_{\text{osc}}$; and main oscillating frequency $\omega_{\text{osc}}$. Their scalings with the model parameters are:
\begin{equation}
\phi_c \sim F, \quad R_{\text{osc}} \sim m^{-1}, \quad M_{\text{osc}} \sim V~R^3_{\text{osc}} \sim F^2/m, \quad \omega_{\text{osc}} \sim m.
\label{eq:oscillon_properties}
\end{equation}
Their precise numerical values slowly depend on time and are model dependent. The goal of this section is to estimate these when the oscillon is on the verge of decaying as well as the oscillon lifetime. Again, we will make a distinction between $p<0$ and $p \geq 0$, picking $p = \pm 1/2$ as representatives.

As it evolves, an oscillon oscillates with a typical main frequency that is very close to but smaller than $m$. During each oscillation it radiates away a very small amount of axion waves, which makes it lose some energy. The remaining energy gets redistributed by a lowering of $\phi_c$ and a slow increase in $R_{\text{osc}}$. This forces the field to oscillate at a slightly larger frequency, but always below $m$. This process of radiation emission and energy redistribution continues until the field is forced to oscillate too close to $m$, at which point all that remains is radiated away in a final burst of axion waves of frequency $\omega \approx m$. Intuitively, this happens when the field reaches values at which its potential is too close to quadratic, \textit{i.e.} when $\phi_c \approx F$.

\begin{figure}[t]
\centering
\includegraphics[width=\textwidth]{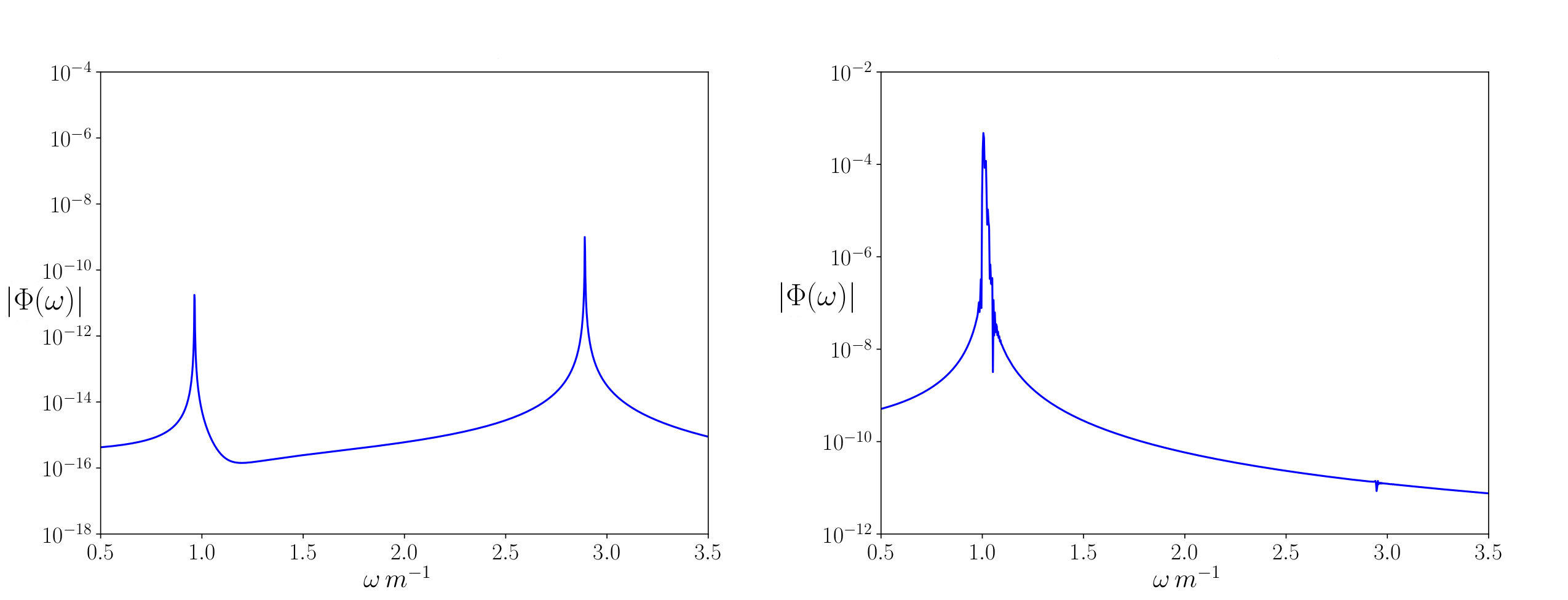}
\caption{\it Power spectrum of the field $\phi$ for $p = 1/2$ at a distance $r = 80 \, m^{-1}$, with arbitrary normalization. Left: Evaluated at $t = 10^7 m^{-1}$. Right: At decay, $t \approx 3 \times 10^8 m^{-1}$. Notice that the peak in the right plot at $\omega\sim m$ is much larger than the peaks in the left plot, meaning that indeed most of the energy of the oscillon is lost in moderately relativistic axion waves at its decay.}
\label{fig:power_spectrum}
\end{figure}

In order to qualitatively elucidate the evolution of an oscillon, the latter can also be seen as a non-perturbative object consisting of a large number $N\sim F^{2}/m^{2}$ of axion quanta \cite{Dvali:2017ruz}. Because the field at the core explores large values, non-conserving particle number processes take place, the most important being $3 \to 1$, and these allow individual axion quanta to escape from the influence of the rest. This picture matches very well the one in \cite{Mukaida:2016hwd,Ibe:2019vyo}, where the extreme longevity of oscillons is understood as arising from a small violation of an effective $U(1)$ symmetry that appears in the non-relativistic limit and corresponding to particle number. This violation is responsible for particle-number violating processes; its smallness implies that such processes are not very frequent and thus that oscillons are very long lived. Numerically, this can be checked by looking at the power spectrum of the waves emitted by an oscillon far away from its core, which we show in Fig.~\ref{fig:power_spectrum}. The dominant peak at $3 \omega_{\text{osc}}$ clearly signals that axions escape through $3 \to 1$ processes. However, the energy radiated through these processes is a negligible fraction of the total oscillon mass. Nonetheless, once too many axions have escaped, the remaining ones cannot generate a large enough collective coupling that self-sustains the system, at which point the oscillon finally decays.

In what follows, we present numerical results of oscillon evolution, assuming spherical symmetry and neglecting Hubble friction. Before doing so, let us therefore comment on the validity of these approximations.

Regarding the assumption of spherical symmetry, it is true that, in general, parametric resonance will not generate perfectly spherically symmetric profiles. However, there is evidence that those perturbations that break spherical symmetry decay quickly \cite{Hindmarsh:2006ur}. Since spherically symmetric oscillons are attractors, we do not see any reason why this result should not be general and model-independent. 

\begin{figure}[t]
\centering
\includegraphics[width=\textwidth]{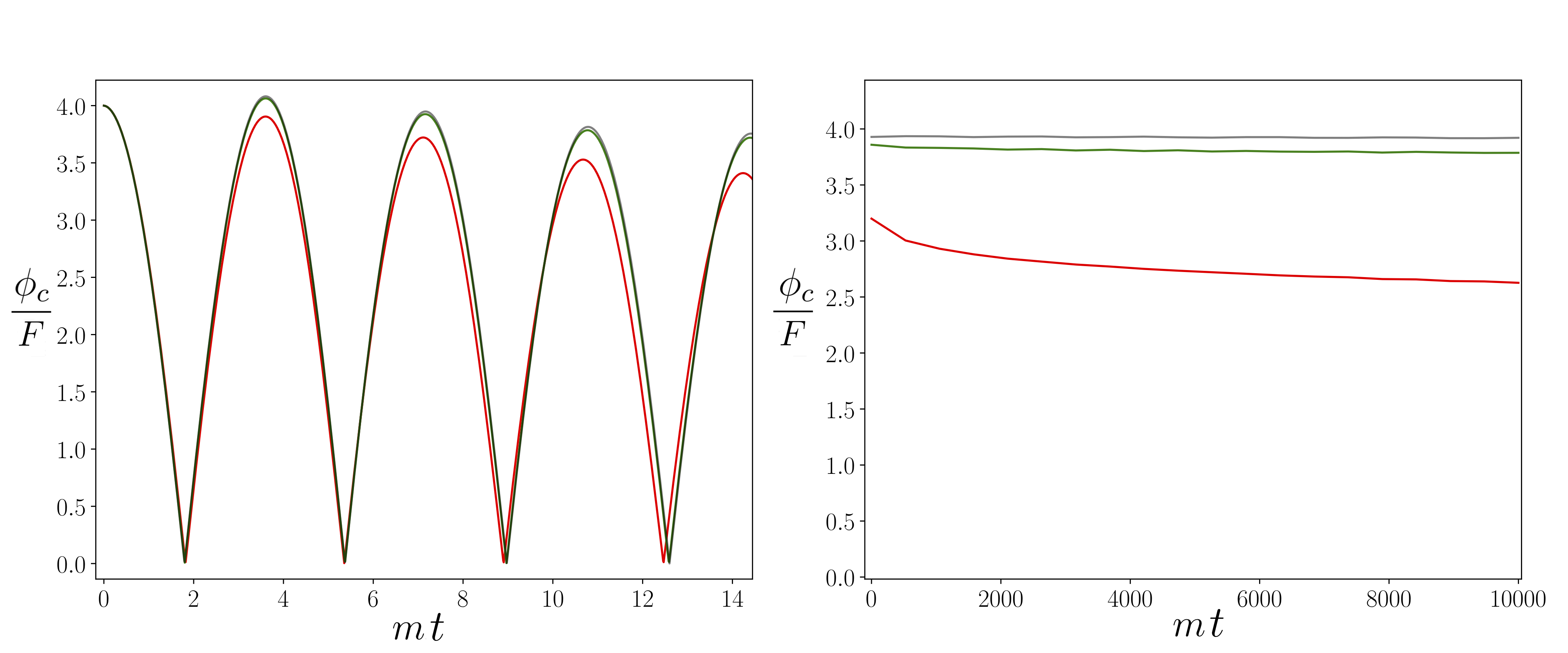}
\caption{\it Effect of Hubble friction with initial values of $H = 10^{-2}, \, 10^{-3}$ (red, green) compared to evolution in Minkowski space (gray). On the left, the first few oscillations; on the right, an average of the oscillations over longer timescales. The plots shown here are for $p = 1/2$. For the expected value of $H\lesssim 10^{-3} \, m$ at oscillon formation, the effect of Hubble friction is negligible.}
\label{fig:hubble_effect}
\end{figure}

As for the consistency of neglecting Hubble friction, we have seen in the previous section that parametric resonance produces significant inhomogeneities after $H \lesssim 10^{-2} \, m$. While we have not performed the full numerical analysis of oscillon formation from parametric resonance, recent simulations with similar models \cite{Lozanov:2017hjm, Kitajima:2018zco} suggest that oscillons actually form a bit later, once $H \lesssim 10^{-3} \, m$~\cite{Kitajima:2018zco}. Since the oscillon has a characteristic size at formation $R_{\text{osc}} \sim m^{-1} \ll H^{-1}$, one expects that the impact of Hubble friction on the evolution of the oscillon should be minimal. In order to support this expectation, in Fig.~\ref{fig:hubble_effect} we have plotted the first few oscillations of a $p = 1/2$ oscillon, including Hubble friction in the equations of motion. It is evident from Fig.~\ref{fig:hubble_effect} that for $H\lesssim 10^{-3}~m$ the effect of friction on the evolution of the oscillon is quite small and we can thus restrict our simulations to Minkowski space.

\begin{figure}[t]
\centering
\includegraphics[width=0.9\textwidth]{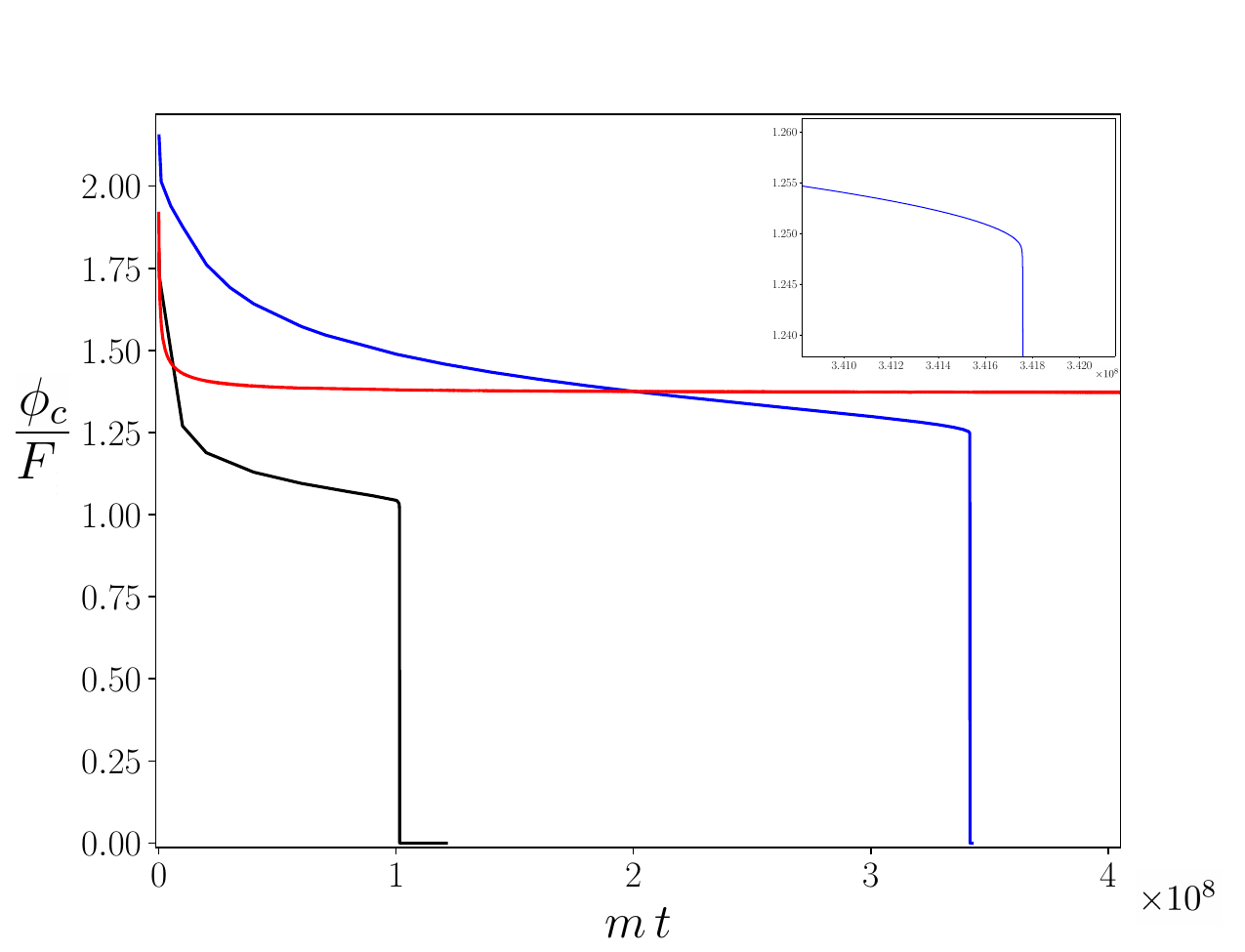}
\caption{\it The life of an oscillon in a snapshot. Here we plot the mean of the absolute value of the field at the center of the oscillon, $\phi_c$, as a function of time for $p = 1/2$ (blue), $p = 0$ (black) and $p = -1/2$ (red).}
\label{fig:oscillon_lifetime}
\end{figure}

Let us now present the results of the numerical evolution of oscillon profiles, according to \eqref{eq:eom}. As in the previous sections, we make a distinction between $p \geq 0$ and $p < 0$, choosing $p = \pm 1/2$ as its representatives. We find that the potential with $p = 1/2$ supports oscillons with a lifetime $\tau \approx 3 \times 10^8 \, m^{-1}$. For $p=-1/2$ we quote only a lower bound $\tau \geq 6.5 \times 10^8 \, m^{-1}$ on the lifetime because we have not seen the oscillons to decay in our simulations after $~1200$ CPU hours. A snapshot of the lifetime of representative oscillons can be seen in Fig.~\ref{fig:oscillon_lifetime}, where we have also included the $p = 0$ case to show that potentials with $0 \leq p <1$ lead to comparable oscillon lifetimes. The values of $A$ and $\sigma$ that we used as initial conditions are $A_{p=1/2} = 2.1~F$, $\sigma_{p=1/2} = 2.45~m^{-1}$; $A_{p=0} = 2~F$, $\sigma_{p=0} = 1.9~m^{-1}$; $A_{p=-1/2} = 2~F$, $\sigma_{p=-1/2} = 1.83~m^{-1}$.

We list the values of $\phi_{c}, R_{\text{osc}}$, $M_{\text{osc}}$ and $\omega_{\text{osc}}$ at the time of oscillon decay for $p = 1/2$ and at $t = 6.5 \times 10^8~m^{-1}$ for $p = -1/2$ in Table \ref{table:critical_oscillon_params}. Numerically we define $R_{\text{osc}}$ as that value of $r$ such that the energy from $r = 0$ to $r = R_{osc}$ is equal to $90 \%$ of the total energy in the lattice.

\begin{table}
\centering
\begin{tabular}{ |c|c|c|c|c|c| }
 \hline
  & $\phi_c/F$ & $M_{osc} \, F^{-2} m$ & $R_{osc} \, m$ & $\omega_{\text{osc}} \, m^{-1}$ & $\tau \, m$ \\
 \hline 
 $p = 1/2$ & $1.25$ & $261.05$ & $9.32$ & 0.98 & $3.4 \times 10^8$ \\
 \hline 
 $p = -1/2$ & $1.37$ & $77.38$ & $5.49$ & 0.93 & $> 6.5 \times 10^8$ \\ 
 \hline
\end{tabular}
\caption{Values of oscillon properties. For $p = 1/2$, these are reported at the time of oscillon decay. For $p = -1/2$, we give the values at $t = 6.5 \times 10^8 m^{-1}$.}
\label{table:critical_oscillon_params}
\end{table}

In conclusion, our analysis shows that oscillons supported by the potentials~\eqref{eq:potmon} have the largest lifetimes ever encountered in the literature. In particular, for $p>0$ we find~$\tau \approx 10^8 \, m^{-1}$. Even more intriguing is the case of $p < 0$, for which we have not yet witnessed oscillon decay after $t = 6.5 \times 10^8 \, m^{-1}$. 


Before moving on to the next chapter of oscillon biography, let us provide details of the numerical method which the results of this subsection are based on. We have evolved an oscillon profile with initial conditions set by \eqref{eq:profile} and $\dot{\phi}(t = 0) = 0$, using the full non-linear equations~\eqref{eq:eom}. We implemented a staggered leapfrog scheme with a three-point spatial Laplacian accurate to $\OO(dx^2)$. We kept a fixed ratio between spatial and time discretization $dr = 5 dt$, and used Neumann boundary conditions at $r = 0$ and second order absorbing boundary conditions at the end of the grid. The typical sizes of the grids we have used vary between $50$ and $80$ units of physical distance, with $dr = 0.05$. While these are the typical values, we also have examples with $dr = 0.01$ and up to $100$ units of physical distance to check stability and robustness of our numerics. Furthermore, we have checked that absorbing boundary conditions remove radiation efficiently from the lattice without introducing significant noise. We used Python to write our code and it has been benchmarked by reproducing the results in \cite{Salmi:2012ta}.

As a final comment, the time resolution, $dt \sim 10^{-2} ~ m^{-1}$, and the current CPU frequencies, $\sim$ GHz,  limit the temporal size of our simulations to at most  $\sim 10^9~m^{-1}$. 
In practice, then, if in a simulation we see the oscillon decay at some time, we take that as a measure of its  lifetime. 
Instead, if in the simulation we do not see its decay, then at most we can place a lower bound on its lifetime.

\subsection{Legacy}
\label{sub:legacy}

Let us now discuss the fate of oscillons once they decay, having in mind a  cosmological setup where they are part of the DM sector. 

Oscillons sustain themselves because of the field self-interactions, which are relevant at large field values and require a relativistic treatment. Localized, spherically symmetric field configurations however arise also in the non-relativistic, small amplitude regime of an oscillating axion field. These are commonly referred to as {\it solitons} and, in contrast with oscillons, are sustained by gravity. Solitons are in fact essential to the proposal of fuzzy dark matter~\cite{Hu:2000ke, Hui:2016ltb} (see also~\cite{Deng:2018jjz, Bar:2018acw} for recent critical takes), since they are supposed to make up the cores of Ultra-Light axion Dark Matter halos. Momentarily neglecting self-interactions, the soliton configuration is obtained as a solution of the non-relativistic Schr\"odinger-Poisson equations \cite{Bar:2018acw}. In fact, a whole family of solutions exist, described by a single parameter $\lambda$. The masses and core radii of these solitons are given by
\begin{align}
\label{eq:msol}
M_{\text{sol}}&=\lambda M_{1},~\text{with}~M_{1}\approx 2.79\times 10^{12}\left(\frac{m}{10^{-22}~\text{eV}}\right)^{-1}M_{\odot}\\
\label{eq:rsol}
R_{\text{sol}}&=\lambda^{-1}R_{1},~\text{with}~R_{1}\approx 0.082~\left(\frac{m}{10^{-22}~\text{eV}}\right)^{-1}~\text{pc},
\end{align}
with the additional requirement $\lambda<1$.
In the absence of initial conditions on $M_{\text{sol}}$ and/or $R_{\text{sol}}$, it is not possible to predict uniquely sizes and masses of axionic gravitational solitons. It is our aim in this section to qualitatively argue that oscillons can indeed work as seeds for the later formation of gravitational solitons, thereby setting the value of $M_{\text{sol}}$ and thus of the parameter $\lambda$.\footnote{The solitons that obey Eqs.~\eqref{eq:msol} and~\eqref{eq:rsol}, are strictly speaking coherent solutions of the Schr\"odinger-Poisson \cite{Bar:2018acw}. 
The most likely outcome of the oscillon decay can be expected to be a non-coherent but gravitationally bound lump, akin to what is usually referred to as miniclusters/minihalos.
It may take a long time for these to settle down to the coherent soliton solution -- we thank S. Sibiryakov for pointing this out to us.
In this discussion we are interested in estimating how the decaying oscillon can be trapped in its own gravitational potential, regardless of its coherence. To this aim we use Eqs.~\eqref{eq:msol} and~\eqref{eq:rsol} only to obtain rough estimates soliton/miniclusters properties. Henceforth we will also refer as soliton both the coherent and incoherent lumps.}

The argument is sharper when interactions among oscillons can be neglected. Let us thus first consider such a simplified scenario. The last stages of the oscillon evolution proceed via the emission of axion waves of momentum $k\simeq R_{\text{osc}}^{-1}\lesssim m$, where in the last step we have neglected a model-dependent (in particular $p$-dependent) numerical prefactor of $O(0.1\div 1)$. Due to Hubble friction, these waves lose energy; eventually they enter the non-relativistic regime and can organize themselves in a soliton structure because of gravity -- falling into their own gravitational potential well. Under our simplifying assumption, the total mass carried away by the waves is equal to the original oscillon mass $M_{\text{osc}}\simeq O(100)\times F^{2}/m$. Intuitively, the formation of a soliton of this mass is possible only if the velocity of the radiated waves is not larger than the escape velocity of the soliton itself. Let us estimate how the escape velocity changes with the scale factor, noticing that the initial radius of the overdensity is given by $R_{\text{osc}}\sim m^{-1}$. The gravitational escape velocity is then given by
\begin{equation}
\label{eq:escape}
v_{e}(t)=\sqrt{\frac{R_{s, \text{osc}}}{R_{\text{osc}} a(t)}}\simeq \frac{1}{\sqrt{a(t)}}\left(\frac{F}{M_{p}}\right),
\end{equation}
where $R_{s, \text{osc}}$ is the Schwarzschild radius of the oscillon and $M_{p}=(8\pi G)^{-1/2}$ is the reduced Planck mass.
The group velocity of the radiated waves is initially moderately relativistic and very rapidly follows the Hubble flow, meaning that it redshifts simply as $v_{q}\sim k/(m a(t))\lesssim 1/a(t)$. In other words, the escape velocity decreases more slowly than the velocity of the radiated waves, which implies that indeed eventually the latter are captured by their own gravitational attraction.\footnote{After decay, the oscillon can be understood as a `wave-packet' of size $\sim (\text{few}) m^{-1}$. Initially, it spreads at a constant speed but one can check that this `peculiar' speed starts feeling the effect of Hubble friction (and decreasing as $1/a(t)$) on a short time scale. The initial deviation from the Hubble flow may imply that not all of the oscillon mass is actually gravitationally captured in a soliton configuration. The estimate $M_{\text{sol}}\simeq M_{\text{osc}}$ should nevertheless capture the right order of magnitude.}
This occurs of course when $R_{\text{osc}}a(t)\simeq R_{\text{sol}}$, as can be easily checked, by taking 
$\lambda=M_{\text{sol}}/M_{1}\approx (F/M_{p})^2$. It is useful to report the corresponding soliton radius, according to \eqref{eq:rsol}
\begin{equation}
\label{eq:rsol2}
R_{\text{sol}}\simeq \frac{c}{m}\left(\frac{M_{p}}{F}\right)^{2},
\end{equation}
where $c\sim 0.1\div 1$ is a model-dependent numerical prefactor.
We thus see that the soliton is parametrically much larger in size than its parent oscillon.

The analysis that we have just performed is relevant in the case where interactions among oscillons are negligible. This is likely the case during the epoch of radiation domination, since oscillons just keep being further separated from each other.
However, during matter domination, overdensities of oscillons grow due to gravity. In this case, gravitational interactions of oscillons can become important: for example, oscillons can merge or disrupt or even lead to the formation of black holes~\cite{Cotner:2018vug} (see Sec.~\ref{sub:after} for further discussion of this possibility). As a result, it is more challenging to estimate the properties of a soliton which forms out of an overdensity of oscillons without numerical simulations. While this detailed understanding goes beyond the aim of this paper, we expect that a soliton will indeed form, possibly of much larger mass. According to \eqref{eq:msol} and \eqref{eq:rsol}, the resulting soliton radius can be much smaller than \eqref{eq:rsol2}. We will discuss in more detail how this uncertainty affects constraints on our scenario in Sec.~\ref{sec:pheno}.

A comment on the effects of self-interactions is now in order. The soliton solution reviewed here is indeed obtained in the non-relativistic regime of a free massive scalar. The effects of the attractive self-interactions dictated by the potentials \eqref{eq:potmon} are negligible as long as they are subdominant compared to the gradient energy of the field. Self-interactions have been properly included in the non-relativistic regime by~\cite{Schiappacasse:2017ham}: the result is that self-interactions may induce a collapse instability of the soliton, if its size is less than 
\begin{equation}
\label{eq:minrad}
R_{\text{min}}\simeq 0.1~\text{pc}\left(\frac{10^{-19}~\text{eV}}{m}\right)\left(\frac{10^{15}~\text{GeV}}{F}\right).
\end{equation}
In most of the parameter space which we will consider in the next section, $R_{\text{sol}}$ is generically quite larger than 1 pc, according to \eqref{eq:rsol2}, thus the corresponding soliton lies in the stable branch, where gravity dominates. However, for $m\gg 10^{-19}\text{eV}$ and $F\gg 10^{15}~\text{GeV}$, the soliton size can be smaller than \eqref{eq:minrad}. Similarly, if multiple oscillons lead to a single soliton, its size can be smaller than~\eqref{eq:rsol2}. In these cases the possibility of collapse driven by self-interactions cannot be entirely excluded.

\section{Observational impact}
\label{sec:pheno}

In the previous sections we have established that oscillons can form from an homogeneous scalar field during the radiation dominated era and survive for very long epochs. We now investigate the observational implications of their presence as part of the DM sector. We aim at providing a list of possible important implications and constraints, while we leave detailed analyses of some of these consequences for future work.

These certainly strongly depend on the fraction $f\equiv \rho_{\text{oscillons}}/\rho_{\text{DM}}$ of scalar dark matter in the form of oscillons. As already mentioned, numerical simulations~(\cite{Amin:2011hj, Liu:2018rrt} in the inflationary context for $p=1/2$ and in~\cite{Kitajima:2018zco} for plateau-like potentials) show that a fraction of $O(1)$ of the energy density of the scalar field resides into oscillons after parametric resonance. However, we expect this conclusion to depend on $p$. In particular, $f$ should decrease as $p$ gets closer to unity, since after all the quadratic potential does not support oscillons. 
In this work, we thus mostly assume that, after parametric resonance, the energy density of the axion field is equally distributed into two components: the original oscillating homogeneous background and a population of massive oscillons. Both components behave as dark matter, albeit of very different mass. 

Implications of our scenario further depend on the typical lifetime of oscillons generated by parametric resonance. In the previous section we argued that attractor configurations exist, with $\tau\gtrsim 10^{8}~m^{-1}$ or larger. While we do not know how many of the oscillons produced by parametric resonance are so long-lived, we provided evidence that the precise initial condition provided by the fragmentation is not expected to strongly affect the longevity of oscillons. Therefore, in what follows we assume that their typical lifetime is indeed given by the numerical results discussed in the previous section.

Oscillons decay away via the radiation of scalar waves. This can in principle occur during radiation or matter domination, depending on the field mass and on the oscillon lifetime. In this section we mostly focus on the case $p=1/2$, which is representative of potentials which support long-lived oscillons with $\tau\sim 10^{8}~m^{-1}$. In Fig.~\ref{fig:opspace}, the thick vertical brown line shows the corresponding value of the ULDM mass for which oscillons decay at matter-radiation (MR) equality, while the dashed brown line corresponds to stability until today. On the upper horizontal axis, values of the oscillon radius in kpc are shown. Solid black lines show the corresponding oscillons masses. In Fig.~\ref{fig:opspace2} we plot values of $M_{\text{osc}}$ and $R_{\text{osc}}$ for larger ULDM masses, for which oscillons decay before MR equality. Different observational constraints arise in the two mass ranges shown in Fig.~\ref{fig:opspace} and in Fig.~\ref{fig:opspace2}. Therefore, we discuss these two scenarios separately. We devote a separate subsection to the observational impact of oscillons which survive across the age of the Universe.

Before moving on however, let us comment on a general constraint which applies in the entire range of scalar field masses which we will be discussing. If we consider a light (pseudo)scalar field which exists already during cosmological inflation, then its fluctuations are of isocurvature type and are thus strongly constrained by CMB observations~\cite{Akrami:2018odb}. The magnitude of these isocurvature fluctuations compared to the scalar adiabatic power spectrum for a light scalar field is simply dictated by the Hubble rate as well as by the value of the field during inflation, which in our case is $\phi_{0}\gtrsim 10~F$. The Hubble rate can be traded for the value of the tensor-to-scalar ratio $r/0.1\simeq 10^{9} (H_{I}/M_{p})^2$, the current upper bound being $r\lesssim 0.07$~\cite{Akrami:2018odb}, while future observations can likely probe values down to $r\gtrsim 10^{-3}$. Therefore, the constraint on isocurvature fluctuations translates into an upper bound on $H_{I}$ or on $r$, if a given value of $F\gtrsim H_{I}$ is specified. It turns out that for $10^{14}~\text{GeV}\lesssim F\lesssim 10^{16}~\text{GeV}$ the ULDM scenario which we focus on in this paper requires very small values of $r$ to evade isocurvature constraints, below the aforementioned observable range.\footnote{The upper bound on $F$ can in principle relaxed if $\phi_{0}\gg 10~F$, however then the homogeneous oscillations overproduce dark matter.} Therefore, if $r$ is eventually measured, the ULDM scenario discussed here, with large $F$, will be in tension with observations (similarly as in~\cite{Hui:2016ltb}).

\subsection{Decay after MR equality}
\label{sub:after}

Let us first focus on the scenario in which oscillons live up to or beyond the epoch of  MR equality. We fix $p=1/2$ as a representative example. This scenario then occurs for $m\lesssim 10^{-19}~\text{eV}$, which is the typical range of so-called ULDM candidates.

\begin{figure}[!t]
\centering
\includegraphics[width=\textwidth]{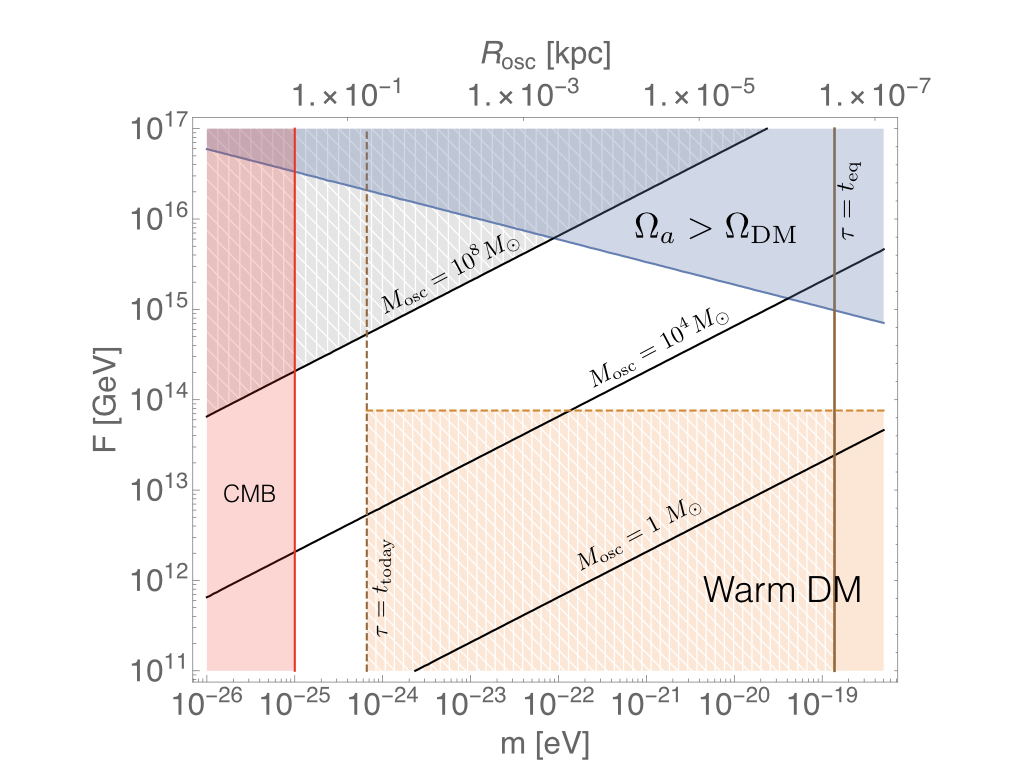}
\caption{\it Properties and constraints of oscillons as functions of $F$ and $m$
in the region where they decay after MR equality.
We take  the representative case $\phi_{0}=15~F$ and $p=1/2$, for which the lifetime is $\tau\simeq 3\cdot10^{8}m^{-1}$.
The red-shaded area is in tension with CMB data and ULDM scalars being most of  the DM \cite{Marsh:2015xka}.
The gray-grid-shaded area corresponds to oscillons with masses above $10^8 M_\odot$, which we take as a rough reference exclusion value. 
In the orange-grid-shaded region the oscillon deaths would warm up DM too much according to the estimate of \eqref{eq:constr1}. 
We didn't include here the constraint $m \gtrsim (10^{-23} - 10^{-21})~\text{eV}$ from \cite{Bozek:2014uqa,Bar:2018acw} because these analyses 
can be affected by the presence of oscillons.}
\label{fig:opspace}
\end{figure}

Several constraints exist for such ultra light scalars; however they are mostly derived assuming that the DM is made only by an oscillating homogeneous field with a cosine potential at MR equality. In particular, \cite{Bozek:2014uqa, Bar:2018acw} suggest that masses below $\sim 10^{-22}~\text{eV}$ are in tension with observations. Interestingly, the presence of oscillons seems to affect significantly these analyses. On the one hand, the constraint from \cite{Bozek:2014uqa} originates from the (otherwise) absence of early galaxies. However, the presence of oscillons means that DM already has large inhomogeneities at MR equality, which are expected to facilitate early structure formation. This suggests that the limit should be revised in this context. On the other hand, the constraint presented in \cite{Bar:2018acw} is based on the simulations of \cite{Schive:2014dra} that evolve the ULDM configuration starting from MR equality. However, again, the presence of oscillons alters significantly the initial condition at MR, so the analysis does not apply in a straightforward manner. For these reasons we do not show these constraints in Fig.~\ref{fig:opspace}.

ULDM masses below $m\sim 10^{-25}~\text{eV}$ are strongly constrained by the primary CMB~\cite{Hlozek:2014lca}, since they alter the expansion rate, thereby affecting the heights of the higher acoustic peaks. This constraint is shown as a shaded red region in Fig~\ref{fig:opspace}. However, in the case of a ULDM with potential \eqref{eq:potmon} and initial value $\phi_{0}>F$, this region is expected to move to larger masses, since ULDM oscillations are delayed compared to the standard axion DM case.

The allowed values of $F$ and $m$ are certainly constrained by dark matter overproduction. For large initial field values, the axion starts oscillating once the Hubble rate drops to $H_{\text{osc}}$ given in \eqref{eq:hosc}.
Shortly afterwards, at $t_{\text{res}}\lesssim 100~m^{-1}$ parametric resonance becomes effective and oscillons are formed. Therefore, the total relic abundance of the axion field can be roughly estimated by redshifting the energy density at $V(\phi_{\text{res}})$ from $t_{\text{res}}$ to today. For the representative initial value $\phi_{0}=15~F$, the result of this calculation is given by the thick blue line in Fig.~\ref{fig:opspace}, the shaded blue region above it being excluded by DM overproduction.

\noindent In our study of the oscillon evolution we have neglected the effects of gravity on the stability of the oscillon itself. This simplifying assumption is certainly valid at early times, when oscillons do not represent a large density contrast compared to the radiation dominated universe. As the Universe cools down to the epoch of MR equality however, gravity can play an important role because of two different effects:
\begin{itemize}
\item[$1$]{An individual oscillon can be seen as an overdensity $\delta\rho_{\text{osc}}$ of dark matter compared to the background homogeneous oscillating field. At formation, the density contrast is
\begin{equation}
\frac{\delta\rho_{\text{oscillon}}}{\rho(t_{\text{osc}})}\simeq \frac{M_{\text{osc}}R_{\text{osc}}^{-3}}{0.1~m^{2}F^{2}}\simeq 1.
\end{equation} 
Afterwards, the field inside the oscillon is essentially decoupled from cosmological expansion, while the background continues to redshift. Thus the density contrast $\delta\rho_{\text{osc}}/\rho_{DM}$ increases with time. Therefore, close to MR equality, oscillons individually start to experience the gravitational force, similarly to what happens in the case of the QCD axion overdensities~\cite{Kolb:1993hw}. Numerical simulations including gravity \cite{Ikeda:2017qev}, albeit in a different context, suggest that oscillons are not destabilized by gravity and can in fact have larger lifetimes. This is at least partially supported by intuition, since gravity acts as another source of attractive interaction.}
\item[$2$]{Oscillons can also be regarded as a dark matter component, independently of the oscillating background from which they originate. They are weakly interacting among themselves and can be very massive, (see the solid lines in Fig.~\ref{fig:opspace}). In this respect, they are similar to MACHOs, though not quite as dense since $R_{\text{osc}}/R_{s, \text{osc}}\sim (M_{p}/F)^{2}$. Due to their low density, they seem to evade rather generically standard constraints on MACHOs and PBHs. However, at MR equality overdensities of oscillons start to grow (as $a(t)$) and gravitational interactions (e.g. merging of oscillons, tidal disruption) may become relevant and again affect oscillon lifetimes (see~\cite{Amin:2019ums} for the effects of gravitational interactions in the non-relativistic regime).}
\end{itemize}
The points above can be concretely addressed via including gravity in the evolution of a single oscillon and by simulating the dynamics of a population of oscillons. We leave these detailed studies for future work and focus here on the case in which oscillons simply decay away after a time $\tau\sim 10^{8} m^{-1}$, by the radiation of axion waves with momentum $k\simeq R_{\text{osc}}^{-1}\lesssim m$.

Further constraints arising from the decay products and legacy of oscillons can restrict the allowed values of $F$ and $m$ in Fig.~\ref{fig:opspace}. However, in the case in which this decay occurs during matter domination, gravitational interactions make precise estimates challenging, as we outlined in Sec.~\ref{sub:legacy}. Therefore, here we simply point out the physical origin of those constraints, while we leave detailed estimates for future work.

Firstly, axion waves radiated from the oscillon decay are in principle warm enough to affect the good behavior of CDM at and after MR equality.
The fraction of energy density in radiated (unbound) axions are constrained to constitute at most few percent of the total DM abundance (see e.g.~\cite{Tanabashi:2018oca} for the similar case of hot DM scenarios). However, as discussed in Sec.~\ref{sub:legacy}, such axions are likely eventually trapped in a soliton configuration. This should occur rapidly enough not to affect structure formation: one possible constraint to impose is that the transition from oscillon to soliton occurs in less than a Hubble time, i.e.~$R_{\text{sol}}\lesssim H^{-1}(\tau)$. As argued in Sec.~\ref{sub:legacy}, it is in fact not clear how large $R_{\text{sol}}$ should be. An over-conservative estimate of $R_{\text{sol}}$ is given by \eqref{eq:rsol2} and gives
\begin{equation}
\label{eq:constr1}
R_{\text{sol}}\lesssim H^{-1}(\tau)\simeq \tau\Rightarrow F\gtrsim \frac{M_{p}}{\sqrt{m\tau}}.
\end{equation}
This constraint is represented in Fig.~\ref{fig:opspace} for $\tau=10^{8}~m^{-1}$ by the grid-shaded orange region and should be taken with a grain of salt. As we will discuss in the next subsection, the constraint becomes more reliable when oscillons decay before MR equality: accordingly, we show it as an orange shaded region (without grid lines) in Fig.~\ref{fig:opspace}.

A different kind of constraint can arise from galactic observations. One the one hand, if oscillons indeed transition to solitons, the size of the latter today cannot be larger than the typical core of DM halos, that is $R_{\text{sol}}\lesssim 1~\text{kpc}$. Again, the uncertainty on $R_{\text{sol}}$ does not allow more detailed estimates and for this reason we do not show this constraint in Fig.~\ref{fig:opspace}. On the other hand, strong constraints exist on very massive compact objects from CMB as well as from a variety of galactic halo observations (see~\cite{Carr:1997cn} and~\cite{Carr:2018rid} for a recent discussion). In particular, the latter very strongly constrain objects with masses above $10^{8}~M_{\odot}$ to make only a very small fraction of DM today. Two caveats do not allow us to straightforwardly apply this constraint to our scenario: First, oscillons are not quite compact objects for $F\ll M_{p}$, which is the region we consider in this work. Secondly, in most of the parameter space shown in Fig.~\ref{fig:opspace} oscillons do not survive until today; rather, they may act as seeds for very massive gravitational solitons, as described above. However, these solitons would then be much less compact than their parent oscillons. Constraints from CMB are derived under similar assumptions on the nature of the massive object. 
Overall, it is not clear how strong the resulting constraints would be as we move away from the dashed brown line and the blue shaded region in Fig.~\ref{fig:opspace} (i.e.~oscillons surviving until today and making up a significant fraction of DM). 
We choose then to simplify these constraints by replacing them with the cut $M\gtrsim 10^{8}~M_{\odot}$ which corresponds to the masses where the halo object dynamical constraint become very stringent.

Overall, we are thus left with a region of parameter space, roughly centered around $F\simeq 10^{15}~\text{GeV}$ and $m\sim 10^{-22}\text{eV}$, where oscillons survive up to and beyond MR equality and can seed the formation of  gravitational solitons and/or of further early structures. Oscillon seeds should thus be considered as part of the initial conditions of numerical simulations of ULDM (see e.g.~\cite{Schive:2014dra}). In particular, such numerical analyses should be initialized with a significantly non-scale invariant power spectrum, peaked around the scale corresponding to the oscillon radii. This in principle provide the opportunity to test our scenario by comparing the results of simulations with observations of DM halos. 

Finally, let us spend some words on the possibility to form black holes from oscillons which decay during matter domination. This mechanism has been recently considered in the context of reheating after inflation~\cite{Cotner:2018vug}. The basic idea starts from the realization that some initial overdensity of oscillons is statistically possible. Such an overdensity would then grow during matter domination, i.e. $\delta\rho_{\text{oscillon}}/\rho_{\text{DM}}\sim a$, and enter the non-linear regime. If this occurs and the overdensity is large enough, a black hole may form. This is more likely to happen the closer the individual oscillons are to being black holes themselves, which corresponds to $F$ close to $M_P$ in our models, in which case the approximation of neglecting gravitational effects is of course not valid.
In our scenario, we can think of this in the language of oscillon-to-soliton transition: in other words, the growth of oscillon overdensities simply corresponds to the fact that several oscillons fuse into a large soliton which approximately captures the total mass of the overdensity. By means of \eqref{eq:msol} and \eqref{eq:rsol}, we can give a rough estimate of how much mass is required so that (a significant part of) the resulting soliton is a black hole. We need
\begin{equation}
\label{eq:bh1}
R_{s, \text{sol}}=\frac{M_{\text{sol}}}{4\pi M_{p}^{2}}\gtrsim R_{\text{sol}}.
\end{equation}
Let us now write $R_{\text{sol}}$ in terms of $M_{\text{sol}}$, according to \eqref{eq:msol} and \eqref{eq:rsol}
\begin{align}
\nonumber R_{\text{sol}}&=0.082~\text{pc}\times \frac{2.8\cdot 10^{12} M_{\odot}}{M_{\text{sol}}}\left(\frac{m}{10^{-22}~\text{eV}}\right)^{-2}\\
\label{eq:bh2}
&\simeq 3\cdot 10^{49} M_{p}^{-1}\times \frac{2.8\cdot 10^{12} M_{\odot}}{M_{\text{sol}}}\left(\frac{m}{10^{-22}~\text{eV}}\right)^{-2},
\end{align}
where in the second step we have expressed length scales in units of the inverse reduced Planck mass. Plugging \eqref{eq:bh2} in the inequality \eqref{eq:bh1}, we obtain
\begin{equation}
\label{eq:bh3}
M_{\text{sol}}\gtrsim 10^{12}M_{\odot} \left(\frac{m}{10^{-22}\text{eV}}\right).
\end{equation}
For $m\lesssim 10^{-19}~\text{eV}$, \eqref{eq:bh3} gives very large would-be black hole masses $M_{\text{sol}}\gtrsim 10^{9}M_{\odot}$. Black holes of these masses are constrained to make up only a small fraction of the DM by different sets of observations, see e.g.~\cite{Carr:2018rid}. Nevertheless notice that they would form after MR equality, therefore the constraints coming from CMB anisotropies do not apply to this case. 
We can now also estimate the number and mass of the oscillons which would be necessary to form a black hole soliton with mass \eqref{eq:bh3}. Taking $m\sim 10^{-19}\text{eV}$, we see in Fig.~\ref{fig:opspace} that close to the solid blue line (where all of the DM comes from the scalar waves and oscillons) the oscillons have masses $M_{\text{osc}}\sim 10^{3}~M_{\odot}$. CMB and other constraints on MACHOs do not apply in this case, since oscillons are very dilute objects. Therefore, approximately $\sim 10^{6}$ oscillons would have to contribute to the formation of the black hole soliton. While we cannot make any further estimate of how likely this scenario is, let us notice that above the orange shaded region in Fig.~\ref{fig:opspace}, the individual oscillon-to-soliton transition takes less than a Hubble time. Therefore, it might not be so implausible to have a very small, but non-negligible fraction of supermassive black holes originating from the growth of oscillon overdensities in our scenario. These estimates are of course very rough and more detailed analyses are required, nevertheless we find this scenario interesting given the potentially important observational impact of such supermassive objects (see e.g.~\cite{Carr:2018rid}).
 
\subsection{Decay before MR equality}
\label{sub:before}

Let us now discuss some peculiarities of the scenario in which oscillons decay before MR equality. Again, we fix $p=1/2$ as a representative example. Thus this scenario occurs for $m>10^{-19}~\text{eV}$. We limit ourselves to the case which is most interesting for dark matter, and thus focus on $m<H_{\text{QCD}}$. 
\begin{figure}[!t]
\centering
\includegraphics[width=.95\textwidth]{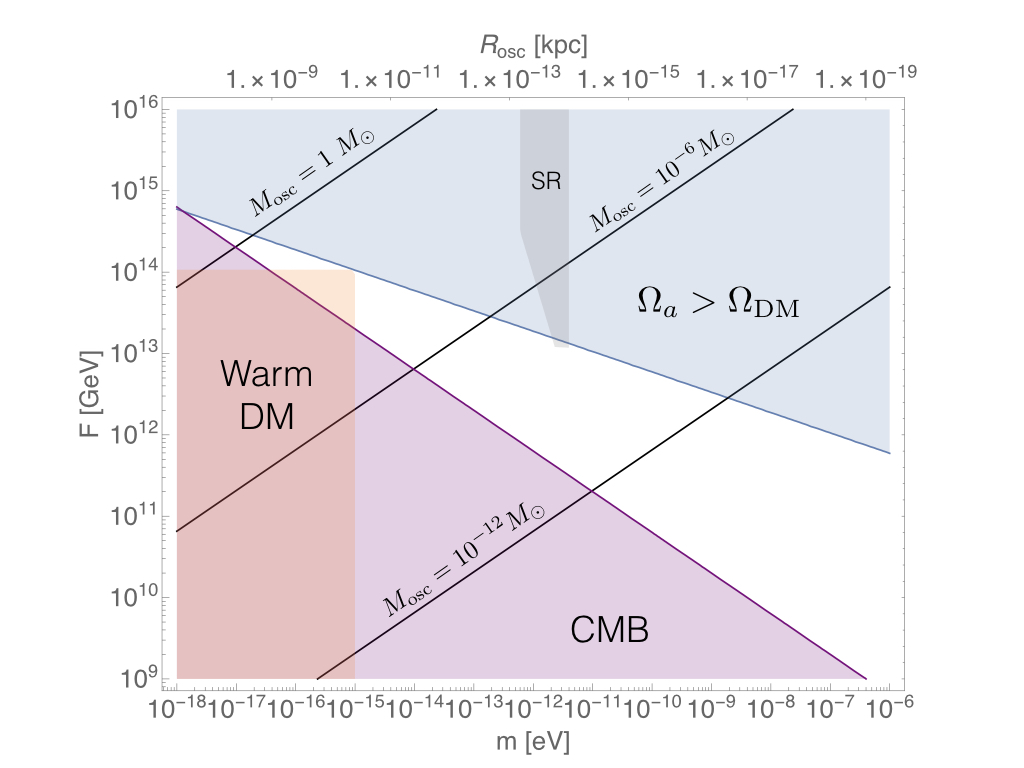}
\caption{\it Properties and constraints of long lived oscillons in the case that they decay before MR equality. As before we take the representative case $p=1/2$ and $\phi_{0}=15~F$, giving $\tau\simeq 3\;10^{8}m^{-1}$. The solid black denote oscillon masses $10^{-12} M_{\odot}, 10^{-6}M_{\odot}$ and $1 M_{\odot}$ from the bottom right to the top left corners. Above the solid blue line  dark matter is  overproduced. The shaded purple region is disfavored by CMB anisotropies and is meant to extend also to the right portion of Fig.~\ref{fig:opspace}, until the brown thick line. The orange shaded region is disfavored by the production of warm dark matter during the decay of the oscillon. The gray shaded region is further constrained by black hole superradiance.}
\label{fig:opspace2}
\end{figure}

In order to match the observed abundance of dark matter, smaller values of $F$ are now required. This is shown in Fig.~\ref{fig:opspace2} by the thick blue line. 
The crucial difference with respect to the previous case is that oscillon overdensities do not significantly grow during radiation domination. After parametric resonance, the typical Hubble patch contains a certain number of separated oscillons. Due to cosmological expansion, the distance between oscillons continues to increase, thus the interactions among them are negligible. Once oscillons decay, their radiated waves are quickly slowed down by Hubble expansion and thus have no possibility to interact with the decay waves of other oscillons. 
Therefore, the size of the resulting soliton should be well estimated by \eqref{eq:rsol2} and we expect the constraint given in \eqref{eq:constr1} to hold. The constrained region is shown in shaded orange in Fig.~\ref{fig:opspace} and Fig.~\ref{fig:opspace2}. Notice however that, for large enough axion masses, the waves radiated by the oscillons are redshifted enough at CMB not to affect structure formation. This occurs for $m\gtrsim 10^{-15}\text{eV}$: these values of axion masses are thus unconstrained by the production of warm dark matter.

A stronger constraint arises however from requiring that the soliton radius is much smaller than the Hubble patch at CMB, i.e.~ $R_{\text{sol}}\ll H^{-1}(t_{\text{eq}})$. In particular, we impose $R_{\text{sol}}< 10^{-3}H^{-1}(t_{\text{eq}})$, given the current resolution of CMB observations. The disfavored region of parameter space is shaded in purple in Fig.~\ref{fig:opspace2}.\footnote{The beginning of this exclusion region is meant to appear in Fig.~\ref{fig:opspace} as well, but given the small parameter space and the size of the uncertainties we dismissed it there.}
The thick lines in the same figure show values of soliton/oscillon masses $1 M_{\odot}, 10^{-6}~M_{\odot}$ and $10^{-12}~M_{\odot}$ from top to bottom.

Yet another source of constraints is the phenomenon of black hole (BH) superradiance \cite{Arvanitaki:2010sy,Arvanitaki:2014wva}, which 
apply even if the (pseudo)scalar is not all of DM. The affected mass ranges are slightly narrow windows around $10^{-12}~\text{eV}$ (from stellar mass BHs), and  $10^{-18}~\text{eV}$ (from supermassive BHs, $M\sim 10^6$), which do not appear in Fig.~\ref{fig:opspace}. Self-interactions play some role in determining the excluded parameter space \cite{Arvanitaki:2010sy,Arvanitaki:2014wva,Arvanitaki:2016qwi,Gruzinov:2016hcq}, which translates into cutting the excluded region for low enough $F$.
Doing so, only the stellar mass BHs bound is marginally stronger than the DM overproduction bound, as shown in Figs.~\ref{fig:opspace2}.
It is natural, however, to ask whether the presence of oscillons can make these bounds more stringent. The oscillon-BH dynamics has been partially studied in previous works. Refs.~\cite{Helfer:2016ljl,Clough:2018exo} show that scalar-gravity systems admit solutions of the form of an oscillon with a BH at the center. Also, Ref.~\cite{Clough:2019jpm} showed that BHs can grow a significant (pseudo)scalar hair. It would be interesting to revise superradiance rates for this kind of BH solutions, but we are not in the position of extracting whether this can have an impact on the superradiance constraints.

The parameter space extends to the mass range of the QCD axion.  Let us thus spend a word on the possibility to have oscillons from the QCD axion potential. In this case, the latter can be computed in the effective chiral lagrangian approach and turns out not to support long lived oscillons. Furthermore, the initial inhomogeneous conditions are provided by the presence of topological defects in the field, if the PQ symmetry is broken after inflation. Thus significant parametric resonance is not required and indeed may be difficult to achieve~\cite{Fukunaga:2019unq}. The formation of oscillons from the annihilation of the QCD axion string-wall network is well known \cite{Kolb:1993hw}. The observational implications are expected to be quite similar to the one we discussed here (see indeed for numerical studies of QCD axitons as seeds of miniclusters~\cite{Vaquero:2018tib, Buschmann:2019icd}).

\subsection{Non-decaying oscillons?}
\label{sec:perpetual_oscillons}

Our phenomenological discussion has focused until now on the case in which oscillons have very long lifetimes, which are however quite short compared to the age of the Universe (unless $m\lesssim 10^{-24}~\text{eV}$, which is however likely to be in tension with observations). We devote this section to the very intriguing possibility that oscillons can survive until today and thus directly make up a significant fraction of the dark matter.

\begin{figure}[!t]
\centering
\includegraphics[width=0.95\textwidth]{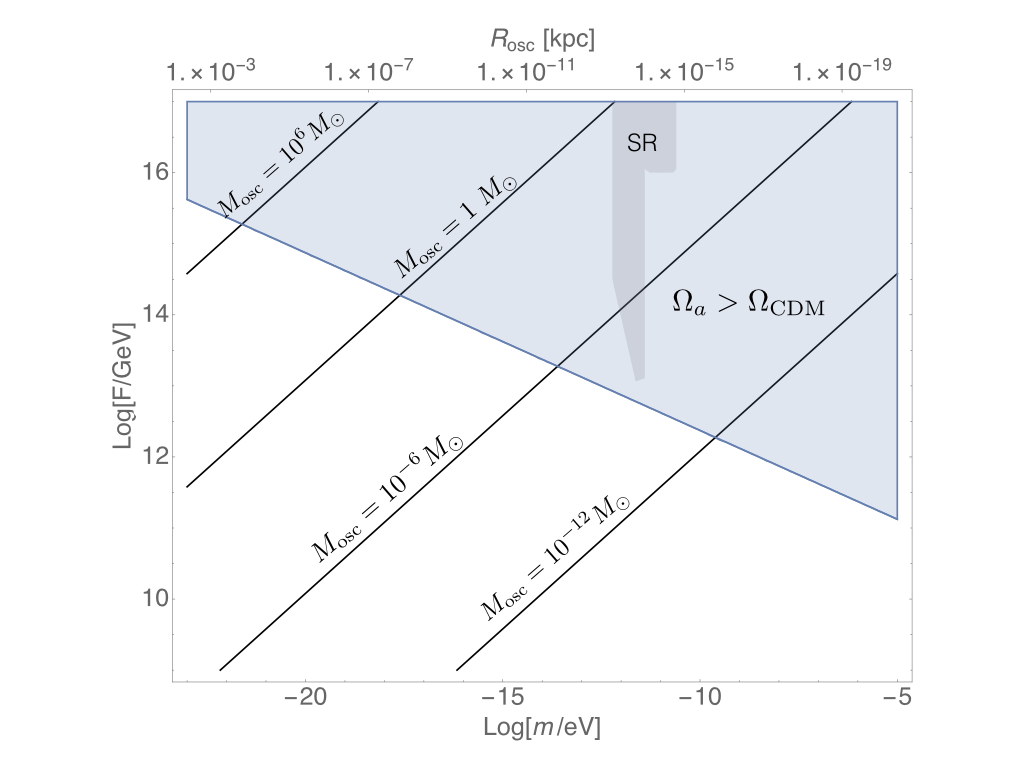}
\caption{\it Properties and constraints of long lived oscillons as functions of $F$ and $m$ assuming that they live up  to today for any $m$ (which might represent changing $p$). Above the solid blue line the homogeneous oscillations of the axion field overproduce dark matter, as extracted from the representative case $p=-1/2$, with $\phi_{0}=15~F$.
The solid black lines denote oscillon masses $10^{6} M_{\odot}, 1 M_{\odot}, 10^{-6}M_{\odot}$ and $10^{-12} M_{\odot}$ from the top left to  bottom right corners. Observationally disfavored masses ($\gtrsim 10^8 M_{\odot}$) are not shown since they lie entirely in the blue shaded region, excluded by DM overproduction. The gray shaded region is further constrained by black hole superradiance~\cite{Arvanitaki:2016qwi}. The tiny mass window around $10^{-18}~\text{eV}$ which is also currently more weakly constrained by supermassive black hole superradiance for $F\gtrsim 10^{16}~\text{GeV}$ is not shown.}
\label{fig:opspace3}
\end{figure}

This speculation is partially supported by two different observations. On the one hand, as mentioned in the previous subsections, the lifetime of oscillons which decay after MR equality may be importantly affected by gravitational dynamics. For example, since oscillon overdensities grow during matter domination, interactions among oscillons are expected to become more relevant in this epoch. In particular, the radiation of some oscillons may end up feeding neighboring oscillons which may then have extended lifetimes. 
Alternatively, oscillons may even coexist with the gravitationally-bound solitons and one can envisage hybrid structures with an oscillon at the cores surrounded by a soliton. One can expect a hybrid object of this sort to be much longer lived because the radiation emitted by the oscillon is trapped inside its own gravitational well, therefore it does not really escape.
Hybrid oscillon-soliton objects have indeed been found numerically in \cite{Ikeda:2017qev}, with infinite lifetimes in some cases.

On the other hand and more importantly for the present work, the lifetime of oscillons crucially depends on the behavior of the axion potential at large field values, i.e. on the exponent $p$. Until now we have focused on the representative case $p=1/2$, for which we know that $\tau\sim 10^{8}~m^{-1}$. The same seems to hold true for any $0\leq p < 1$. However, the situation appears to be different for $p<0$,  i.e. for plateau-like potentials. As discussed in Sec.~\ref{sec:lifetime}, we only have a lower bound on the lifetime for the case $p=-1/2$, i.e. $\tau\gtrsim 6.5~\cdot 10^{8}~m^{-1}$. It is not inconceivable that $\tau_{(p<0)}\gg 10^8~m^{-1}$. In particular, notice that for $m\lesssim 10^{-22}~\text{eV}$, `just' $\tau_{(p<0)}\gtrsim 7\cdot 10^{10}~m^{-1}$ is required to have oscillon stability until the current age of the Universe.

With these motivations in mind, let us consider here the scenario in which oscillons are stable until today for a wide range of phenomenologically interesting axion masses, say  $10^{-23}~\text{eV}\lesssim m\lesssim 10^{-5}~\text{eV}$. In this case, there are essentially no constraints coming from the oscillon decay into warm axions, therefore the available parameter space opens up significantly, as shown in Fig.~\ref{fig:opspace3}. In drawing Fig.~\ref{fig:opspace3}, we are essentially assuming that $\tau\gtrsim t_{\text{today}}$ for every value of $m$. To be more concrete, the required lifetime in the mass range shown in Fig.~\ref{fig:opspace3} is $10^{10}~ \lesssim (\tau \,m) \lesssim 10^{26}$. While this huge range seems implausible given the much shorter lifetime in the case $p=1/2$, preliminary numerical results based on a simple extrapolation of numerical data for the case $p=-1/2$ suggest that such long lifetimes may be not so far-fetched~\cite{inpreparation}.

The region where dark matter is overproduced is shown in Fig.~\ref{fig:opspace3} fixing $p=-1/2$ and $\phi_{0}=15~F$. Its shape does not strongly depend on $p$, as long as $p<0$. For $p>0$, constraints from DM are less rigid, as shown in Figs.~\ref{fig:opspace} and \ref{fig:opspace2}. The oscillon mass as well as its radius are also obtained for $p=-1/2$ and are slightly smaller than in the case $p=1/2$, according to Table~\ref{table:critical_oscillon_params}.

For the masses considered in Fig.~\ref{fig:opspace3}, oscillons always remain smaller than approximately $1~\text{pc}$, therefore they are unconstrained by CMB anisotropies. We should remark that Fig.~\ref{fig:opspace3} is obtained neglecting gravitational interactions among oscillons, which may lead to mergers and solitons as in the previous subsections. Black holes could also form from the growth of oscillon overdensities, as we discussed in the last paragraphs of Sec.~\ref{sub:after}. Interestingly, in the case considered in this section, masses smaller than $10^{-19}~\text{eV}$ are allowed which could lead to lighter and observationally more viable black holes, according to \eqref{eq:bh3}.

While we leave more detailed estimates of oscillon lifetimes for $p<0$ for future work, we cannot stop ourselves from pointing out that for a wide range of decay constants ALP dark matter may be today characterized by a large number of stable oscillons in our Hubble patch.

\section{Conclusions}
\label{sec:conclusions}

We have investigated oscillons (long-lived and localized  scalar field lumps sustained by self-interactions only) and their implications for cosmology. 
Our focus has been on very long-lived oscillons because they arise in certain well-motivated models, and they give a significant impact on dark matter.

Oscillons are known to arise in self-interacting massive scalar field theories whenever the potential $V(\phi)$ exhibits regions which are shallower than quadratic.
We find very long-lived oscillons whenever $V(\phi)$ flattens out at large field values\footnote{A more quantitative condition on the potential for very long lived oscillons seems to be that the effective mass $V''(\phi)$ is not very negative anywhere in the field space. This can happen whenever  i) $V''(\phi)>0$ for all $\phi$, or ii) in case that $V''(\phi)<0$ somewhere,  $|V''(\phi)|/V''(0)\ll1$. One of these conditions is satisfied the benchmark models \eqref{eq:potmon}.}, with lifetimes reaching out to $\tau\sim (10^8 - 10^9)\, m^{-1}$ or more. Benchmark possibilities are a linear or plateau behavior beyond a critical field value $F$. Potentials with these properties are known to arise for axion fields in compactifications of string theory and in supergravity.

Another grace of these models is that they have a built-in {\it oscillogenesis} mechanism: an initial homogeneous field configuration with moderately large field value is enough to end up generating oscillons. The reason is that the homogeneous configuration suffers from parametric resonant instability once the scalar starts to  (homogeneously) oscillate: field fluctuations of certain wavelengths can grow exponentially and thus lead to the fragmentation of the scalar field. This mechanism in turn provides the right ground on which oscillons can form (as confirmed by numerical simulations, see~\cite{Kitajima:2018zco, Fukunaga:2019unq}). We have studied the condition under which parametric resonance is efficient and we find that the localized overdensities are indeed generated for mildly large initial values of the scalar field, not far from $F$.

We then numerically evolved a single oscillon configuration, according to the full non-linear equation of motion, assuming spherical symmetry (and ignoring gravity). We provided numerical evidence that oscillons are essentially unaffected by Hubble friction, therefore we performed most of our simulations in flat space. While we confirm that oscillons radiate scalar waves during their evolution, the loss of energy due to this process can be extremely slow and thus makes the objects long-lived on cosmological scales. In particular, for potentials which go linearly at large field values, or as positive power law with exponent smaller than one, we found that oscillons live up to $10^8$ oscillations (in units of the inverse scalar field mass $m^{-1}$). These are the longest lived oscillons ever encountered to our knowledge. Even more dramatic is the case of plateau-like potentials. In this case we have not yet witnessed oscillon death and so we are only able to put a lower bound on their lifetime,  $\tau \gtrsim 10^9~m^{-1}$.

Such long lifetimes  make oscillons  interesting in the framework of scalar dark matter. In the observationally viable range of $m$, oscillons are always born during the radiation domination epoch from the oscillations of the (pseudo)scalar field. After formation, these objects can be seen as nothing else than bound states of scalar particles, and thus behave as a separate MACHO-like (albeit much less dense) dark matter component, with masses ranging from $10^{-12}$ to $10^{8} M_{\odot}$ depending on $F$ and $m$. The physical radius of these objects at formation ranges from asteroid to galactic-core size and remains essentially constant during the oscillon lifetime, as the object is essentially decoupled from the Hubble flow.  

We investigated several observationally interesting implications of oscillons in the context of scalar ULDM. For one thing, if their lifetimes are shorter than the age of the Universe, oscillons can decay before or after matter-radiation (MR) equality. The impact is  rather sensitive to model details. Assuming $\tau m \sim 10^8$ as a benchmark value for oscillon lifetimes, then $m\leq 10^{-19}~\text{eV}$ corresponds to oscillons death happening at or after MR equality. 
For $m>10^{-19}~\text{eV}$ and $\tau m \sim 10^8$, oscillons decay before MR equality. 
Since at decay oscillons emit a significant amount of warm scalar radiation, this poses some threat to the ULDM model. 
In this respect, gravity comes to rescue because a decaying oscillon can after all be bound by its own gravitational potential, remaining kept as what is now referred to as a soliton (gravitationally-bound scalar field lump, see e.g.~\cite{Schiappacasse:2017ham, Hertzberg:2018lmt, Deng:2018jjz, Bar:2018acw}). 
It would be interesting to further study or simulate numerically the oscillon-to-soliton transition, however this (perhaps simplified) picture already allows to estimate the observational  bounds.

For $m<10^{-19}~\text{eV}$, oscillons would decay in the matter dominated era, when gravitational effects play an important role in the full evolution of the dark sector itself, therefore it is more challenging to extract clean results. While we have not performed simulations including gravity (see~\cite{Amin:2019ums} for simulations of non-relativistic ULDM solitons during matter domination), we can speculate that the latter will actually help the stability of oscillons and extend their lifetimes. Indeed, gravity on the one hand can lead to mergers of oscillons, with the possibility of forming heavier, thus longer-lived objects; on the other hand it causes the growth of overdensities, including in the vicinity of the lumps, thereby providing a mechanism to feed the oscillons with the part of the DM that is in the form of scalar waves. Nevertheless, gravity could also induce destructive effects, such as tidal disruption.

Of course the story can be much richer because gravitationally-bound solitonic structures enter into the game.  
For instance, hybrid objects could form, such as gravitational solitons with oscillon cores~\cite{Ikeda:2017qev} that, perhaps, are longer lived than what we can estimate with flat space simulations.
While this analysis goes well beyond the scope of this paper, it seems that rather robustly the presence of oscillons (in the past or until  today) must be imprinted in the spectrum of DM overdensities as a peak in its spectrum of overdensities at a scale corresponding to the oscillon size, which is roughly the inverse scalar mass $\sim m^{-1}$. 
This would then alter the initial conditions that need to be assumed for numerical simulations of the ULDM such as the one in \cite{Schive:2014dra}.

Finally, we cannot resist entertaining the option that oscillons are stable on the scale of the age of the Universe, which might be the case for plateau-like potentials. 
This scenario would have particularly exciting consequences: namely in this case a significant fraction of the DM today would be in the form of ancient compact objects, similar in a sense to boson stars and solitons, albeit sustained mainly by self interactions. 
A possible signature would be in pulsar timing arrays observatories, which can be sensitive to ULDM models as described in \cite{Khmelnitsky:2013lxt}. The signal is enhanced for denser DM sources, so a larger signal is expected from the galactic centre \cite{DeMartino:2017qsa}. 
In case oscillons live up to today and make up a significant fraction of the DM, nearby oscillons could also lead individually to a sizeable signal. 
It is also interesting to consider the case of multiple species of axions with different masses and decay constants (i.e.~the~\emph{axiverse}). The dark matter today could then be simply made up of oscillons of different sizes and masses, in striking similarity to the organization of baryonic structures around us.
We presented here some basic constraints on these possibilities, while we leave the important and promising task of figuring out the ultimate fate of these oscillons for future work.

\medskip 
\section*{Acknowledgments}
We thank R.Z.~Ferreira, F.~Ferrer, E.~Masso and S.~Sibiryakov for reading a preliminary draft of this paper and for discussions. We thank F.~Ferrer also for helping us in setting up the numerical methods used in this paper. We also thank G.~Dvali and D.~Gaggero for discussions. We thank Manuel Kreutle for useful earlier collaboration. This work was partly supported by the grants FPA2017-88915-P and SEV-2016-0588 from MINECO and 2017-SGR-1069 from DURSI.

\providecommand{\href}[2]{#2}\begingroup\raggedright\endgroup


\begin{thebibliography}{10}

\bibitem{Turner:1983he}
M.~S. Turner, {\it {Coherent Scalar Field Oscillations in an Expanding
  Universe}},  {\em Phys. Rev.} {\bf D28} (1983) 1243.

\bibitem{Press:1989id}
W.~H. Press, B.~S. Ryden, and D.~N. Spergel, {\it {Single Mechanism for
  Generating Large Scale Structure and Providing Dark Missing Matter}},  {\em
  Phys. Rev. Lett.} {\bf 64} (1990) 1084.

\bibitem{Sin:1992bg}
S.-J. Sin, {\it {Late time cosmological phase transition and galactic halo as
  Bose liquid}},  {\em Phys. Rev.} {\bf D50} (1994) 3650--3654,
  [\href{http://arxiv.org/abs/hep-ph/9205208}{{\tt hep-ph/9205208}}].

\bibitem{Hu:2000ke}
W.~Hu, R.~Barkana, and A.~Gruzinov, {\it {Cold and fuzzy dark matter}},  {\em
  Phys. Rev. Lett.} {\bf 85} (2000) 1158--1161,
  [\href{http://arxiv.org/abs/astro-ph/0003365}{{\tt astro-ph/0003365}}].

\bibitem{Goodman:2000tg}
J.~Goodman, {\it {Repulsive dark matter}},  {\em New Astron.} {\bf 5} (2000)
  103, [\href{http://arxiv.org/abs/astro-ph/0003018}{{\tt astro-ph/0003018}}].

\bibitem{Peebles:2000yy}
P.~J.~E. Peebles, {\it {Fluid dark matter}},  {\em Astrophys. J.} {\bf 534}
  (2000) L127, [\href{http://arxiv.org/abs/astro-ph/0002495}{{\tt
  astro-ph/0002495}}].

\bibitem{Amendola:2005ad}
L.~Amendola and R.~Barbieri, {\it {Dark matter from an ultra-light
  pseudo-Goldsone-boson}},  {\em Phys. Lett.} {\bf B642} (2006) 192--196,
  [\href{http://arxiv.org/abs/hep-ph/0509257}{{\tt hep-ph/0509257}}].

\bibitem{Schive:2014dra}
H.-Y. Schive, T.~Chiueh, and T.~Broadhurst, {\it {Cosmic Structure as the
  Quantum Interference of a Coherent Dark Wave}},  {\em Nature Phys.} {\bf 10}
  (2014) 496--499, [\href{http://arxiv.org/abs/1406.6586}{{\tt
  arXiv:1406.6586}}].

\bibitem{Hui:2016ltb}
L.~Hui, J.~P. Ostriker, S.~Tremaine, and E.~Witten, {\it {Ultralight scalars as
  cosmological dark matter}},  {\em Phys. Rev.} {\bf D95} (2017), no.~4 043541,
  [\href{http://arxiv.org/abs/1610.08297}{{\tt arXiv:1610.08297}}].

\bibitem{Bogolyubsky:1976yu}
I.~L. Bogolyubsky and V.~G. Makhankov, {\it {Lifetime of Pulsating Solitons in
  Some Classical Models}},  {\em Pisma Zh. Eksp. Teor. Fiz.} {\bf 24} (1976)
  15--18.

\bibitem{Gleiser:1993pt}
M.~Gleiser, {\it {Pseudostable bubbles}},  {\em Phys. Rev.} {\bf D49} (1994)
  2978--2981, [\href{http://arxiv.org/abs/hep-ph/9308279}{{\tt
  hep-ph/9308279}}].

\bibitem{Copeland:1995fq}
E.~J. Copeland, M.~Gleiser, and H.~R. Muller, {\it {Oscillons: Resonant
  configurations during bubble collapse}},  {\em Phys. Rev.} {\bf D52} (1995)
  1920--1933, [\href{http://arxiv.org/abs/hep-ph/9503217}{{\tt
  hep-ph/9503217}}].

\bibitem{Coleman:1985ki}
S.~R. Coleman, {\it {Q Balls}},  {\em Nucl. Phys.} {\bf B262} (1985) 263.
  [Erratum: Nucl. Phys.B269,744(1986)].

\bibitem{Kusenko:1997si}
A.~Kusenko and M.~E. Shaposhnikov, {\it {Supersymmetric Q balls as dark
  matter}},  {\em Phys. Lett.} {\bf B418} (1998) 46--54,
  [\href{http://arxiv.org/abs/hep-ph/9709492}{{\tt hep-ph/9709492}}].

\bibitem{Honda:2001xg}
E.~P. Honda and M.~W. Choptuik, {\it {Fine structure of oscillons in the
  spherically symmetric phi**4 Klein-Gordon model}},  {\em Phys. Rev.} {\bf
  D65} (2002) 084037, [\href{http://arxiv.org/abs/hep-ph/0110065}{{\tt
  hep-ph/0110065}}].

\bibitem{Kasuya:2002zs}
S.~Kasuya, M.~Kawasaki, and F.~Takahashi, {\it {I-balls}},  {\em Phys. Lett.}
  {\bf B559} (2003) 99--106, [\href{http://arxiv.org/abs/hep-ph/0209358}{{\tt
  hep-ph/0209358}}].

\bibitem{Fodor:2006zs}
G.~Fodor, P.~Forgacs, P.~Grandclement, and I.~Racz, {\it {Oscillons and
  Quasi-breathers in the phi**4 Klein-Gordon model}},  {\em Phys. Rev.} {\bf
  D74} (2006) 124003, [\href{http://arxiv.org/abs/hep-th/0609023}{{\tt
  hep-th/0609023}}].

\bibitem{Saffin:2006yk}
P.~M. Saffin and A.~Tranberg, {\it {Oscillons and quasi-breathers in D+1
  dimensions}},  {\em JHEP} {\bf 01} (2007) 030,
  [\href{http://arxiv.org/abs/hep-th/0610191}{{\tt hep-th/0610191}}].

\bibitem{Hindmarsh:2007jb}
M.~Hindmarsh and P.~Salmi, {\it {Oscillons and domain walls}},  {\em Phys.
  Rev.} {\bf D77} (2008) 105025, [\href{http://arxiv.org/abs/0712.0614}{{\tt
  arXiv:0712.0614}}].

\bibitem{Fodor:2008es}
G.~Fodor, P.~Forgacs, Z.~Horvath, and A.~Lukacs, {\it {Small amplitude
  quasi-breathers and oscillons}},  {\em Phys. Rev.} {\bf D78} (2008) 025003,
  [\href{http://arxiv.org/abs/0802.3525}{{\tt arXiv:0802.3525}}].

\bibitem{Gleiser:2008ty}
M.~Gleiser and D.~Sicilia, {\it {Analytical Characterization of Oscillon Energy
  and Lifetime}},  {\em Phys. Rev. Lett.} {\bf 101} (2008) 011602,
  [\href{http://arxiv.org/abs/0804.0791}{{\tt arXiv:0804.0791}}].

\bibitem{Fodor:2009kf}
G.~Fodor, P.~Forgacs, Z.~Horvath, and M.~Mezei, {\it {Radiation of scalar
  oscillons in 2 and 3 dimensions}},  {\em Phys. Lett.} {\bf B674} (2009)
  319--324, [\href{http://arxiv.org/abs/0903.0953}{{\tt arXiv:0903.0953}}].

\bibitem{Amin:2010jq}
M.~A. Amin and D.~Shirokoff, {\it {Flat-top oscillons in an expanding
  universe}},  {\em Phys. Rev.} {\bf D81} (2010) 085045,
  [\href{http://arxiv.org/abs/1002.3380}{{\tt arXiv:1002.3380}}].

\bibitem{Hertzberg:2010yz}
M.~P. Hertzberg, {\it {Quantum Radiation of Oscillons}},  {\em Phys. Rev.} {\bf
  D82} (2010) 045022, [\href{http://arxiv.org/abs/1003.3459}{{\tt
  arXiv:1003.3459}}].

\bibitem{Amin:2011hj}
M.~A. Amin, R.~Easther, H.~Finkel, R.~Flauger, and M.~P. Hertzberg, {\it
  {Oscillons After Inflation}},  {\em Phys. Rev. Lett.} {\bf 108} (2012)
  241302, [\href{http://arxiv.org/abs/1106.3335}{{\tt arXiv:1106.3335}}].

\bibitem{Salmi:2012ta}
P.~Salmi and M.~Hindmarsh, {\it {Radiation and Relaxation of Oscillons}},  {\em
  Phys. Rev.} {\bf D85} (2012) 085033,
  [\href{http://arxiv.org/abs/1201.1934}{{\tt arXiv:1201.1934}}].

\bibitem{Andersen:2012wg}
E.~A. Andersen and A.~Tranberg, {\it {Four results on phi**4 oscillons in D+1
  dimensions}},  {\em JHEP} {\bf 12} (2012) 016,
  [\href{http://arxiv.org/abs/1210.2227}{{\tt arXiv:1210.2227}}].

\bibitem{Saffin:2014yka}
P.~M. Saffin, P.~Tognarelli, and A.~Tranberg, {\it {Oscillon Lifetime in the
  Presence of Quantum Fluctuations}},  {\em JHEP} {\bf 08} (2014) 125,
  [\href{http://arxiv.org/abs/1401.6168}{{\tt arXiv:1401.6168}}].

\bibitem{Mukaida:2016hwd}
K.~Mukaida, M.~Takimoto, and M.~Yamada, {\it {On Longevity of
  I-ball/Oscillon}},  {\em JHEP} {\bf 03} (2017) 122,
  [\href{http://arxiv.org/abs/1612.07750}{{\tt arXiv:1612.07750}}].

\bibitem{Ibe:2019vyo}
M.~Ibe, M.~Kawasaki, W.~Nakano, and E.~Sonomoto, {\it {Decay of I-ball/Oscillon
  in Classical Field Theory}},  \href{http://arxiv.org/abs/1901.06130}{{\tt
  arXiv:1901.06130}}.

\bibitem{Gleiser:2019rvw}
M.~Gleiser and M.~Krackow, {\it {Resonant Configurations in Scalar Field
  Theories: Can (Some) Oscillons Live Forever?}},
  \href{http://arxiv.org/abs/1906.04070}{{\tt arXiv:1906.04070}}.
  
\bibitem{Dymnikova:2000dy}
  I.~Dymnikova, L.~Koziel, M.~Khlopov and S.~Rubin,
  {\it {Quasilumps from first order phase transitions}},
  {\em Grav.\ Cosmol.}  {\bf 6} (2000) 311,
  [\href{http://arxiv.org/abs/hep-th/0010120}{{\tt hep-th/0010120}}].


\bibitem{Kolb:1993hw}
E.~W. Kolb and I.~I. Tkachev, {\it {Nonlinear axion dynamics and formation of
  cosmological pseudosolitons}},  {\em Phys. Rev.} {\bf D49} (1994) 5040--5051,
  [\href{http://arxiv.org/abs/astro-ph/9311037}{{\tt astro-ph/9311037}}].

\bibitem{Kolb:1994fi}
E.~W. Kolb and I.~I. Tkachev, {\it {Large amplitude isothermal fluctuations and
  high density dark matter clumps}},  {\em Phys. Rev.} {\bf D50} (1994)
  769--773, [\href{http://arxiv.org/abs/astro-ph/9403011}{{\tt
  astro-ph/9403011}}].

\bibitem{Hogan:1988mp}
C.~J. Hogan and M.~J. Rees, {\it {AXION MINICLUSTERS}},  {\em Phys. Lett.} {\bf
  B205} (1988) 228--230.
  

\bibitem{Visinelli:2017ooc}
  L.~Visinelli, S.~Baum, J.~Redondo, K.~Freese and F.~Wilczek,
  {\it {Dilute and dense axion stars}}, 
  {\em Phys. Lett.} {\bf B777} (2018) 64,
   [\href{http://arxiv.org/abs/1710.08910}{\tt arXiv:1710.08910}].


\bibitem{Vaquero:2018tib}
A.~Vaquero, J.~Redondo, and J.~Stadler, {\it {Early seeds of of axion
  miniclusters}},  \href{http://arxiv.org/abs/1809.09241}{{\tt
  arXiv:1809.09241}}.

\bibitem{Buschmann:2019icd}
M.~Buschmann, J.~W. Foster, and B.~R. Safdi, {\it {Early-Universe Simulations
  of the Cosmological Axion}},  \href{http://arxiv.org/abs/1906.00967}{{\tt
  arXiv:1906.00967}}.

\bibitem{Silverstein:2008sg}
E.~Silverstein and A.~Westphal, {\it {Monodromy in the CMB: Gravity Waves and
  String Inflation}},  {\em Phys. Rev.} {\bf D78} (2008) 106003,
  [\href{http://arxiv.org/abs/0803.3085}{{\tt arXiv:0803.3085}}].

\bibitem{McAllister:2008hb}
L.~McAllister, E.~Silverstein, and A.~Westphal, {\it {Gravity Waves and Linear
  Inflation from Axion Monodromy}},  {\em Phys. Rev.} {\bf D82} (2010) 046003,
  [\href{http://arxiv.org/abs/0808.0706}{{\tt arXiv:0808.0706}}].

\bibitem{Dong:2010in}
X.~Dong, B.~Horn, E.~Silverstein, and A.~Westphal, {\it {Simple exercises to
  flatten your potential}},  {\em Phys. Rev.} {\bf D84} (2011) 026011,
  [\href{http://arxiv.org/abs/1011.4521}{{\tt arXiv:1011.4521}}].

\bibitem{Lozanov:2017hjm}
K.~D. Lozanov and M.~A. Amin, {\it {Self-resonance after inflation: oscillons,
  transients and radiation domination}},  {\em Phys. Rev.} {\bf D97} (2018),
  no.~2 023533, [\href{http://arxiv.org/abs/1710.06851}{{\tt
  arXiv:1710.06851}}].

\bibitem{Kitajima:2018zco}
N.~Kitajima, J.~Soda, and Y.~Urakawa, {\it {Gravitational wave forest from
  string axiverse}},  {\em JCAP} {\bf 1810} (2018), no.~10 008,
  [\href{http://arxiv.org/abs/1807.07037}{{\tt arXiv:1807.07037}}].

\bibitem{Schiappacasse:2017ham}
E.~D. Schiappacasse and M.~P. Hertzberg, {\it {Analysis of Dark Matter Axion
  Clumps with Spherical Symmetry}},  {\em JCAP} {\bf 1801} (2018) 037,
  [\href{http://arxiv.org/abs/1710.04729}{{\tt arXiv:1710.04729}}]. [Erratum:
  JCAP1803,no.03,E01(2018)].

\bibitem{Hertzberg:2018lmt}
M.~P. Hertzberg and E.~D. Schiappacasse, {\it {Scalar dark matter clumps with
  angular momentum}},  {\em JCAP} {\bf 1808} (2018), no.~08 028,
  [\href{http://arxiv.org/abs/1804.07255}{{\tt arXiv:1804.07255}}].

\bibitem{Deng:2018jjz}
H.~Deng, M.~P. Hertzberg, M.~H. Namjoo, and A.~Masoumi, {\it {Can Light Dark
  Matter Solve the Core-Cusp Problem?}},  {\em Phys. Rev.} {\bf D98} (2018),
  no.~2 023513, [\href{http://arxiv.org/abs/1804.05921}{{\tt
  arXiv:1804.05921}}].

\bibitem{Bar:2018acw}
N.~Bar, D.~Blas, K.~Blum, and S.~Sibiryakov, {\it {Galactic rotation curves
  versus ultralight dark matter: Implications of the soliton-host halo
  relation}},  {\em Phys. Rev.} {\bf D98} (2018), no.~8 083027,
  [\href{http://arxiv.org/abs/1805.00122}{{\tt arXiv:1805.00122}}].

\bibitem{Marsh:2015xka}
D.~J.~E. Marsh, {\it {Axion Cosmology}},  {\em Phys. Rept.} {\bf 643} (2016)
  1--79, [\href{http://arxiv.org/abs/1510.07633}{{\tt arXiv:1510.07633}}].

\bibitem{Witten:1980sp}
E.~Witten, {\it {Large N Chiral Dynamics}},  {\em Annals Phys.} {\bf 128}
  (1980) 363.

\bibitem{Witten:1998uka}
E.~Witten, {\it {Theta dependence in the large N limit of four-dimensional
  gauge theories}},  {\em Phys. Rev. Lett.} {\bf 81} (1998) 2862--2865,
  [\href{http://arxiv.org/abs/hep-th/9807109}{{\tt hep-th/9807109}}].

\bibitem{Dvali:2005an}
G.~Dvali, {\it {Three-form gauging of axion symmetries and gravity}},
  \href{http://arxiv.org/abs/hep-th/0507215}{{\tt hep-th/0507215}}.

\bibitem{Kaloper:2008fb}
N.~Kaloper and L.~Sorbo, {\it {A Natural Framework for Chaotic Inflation}},
  {\em Phys. Rev. Lett.} {\bf 102} (2009) 121301,
  [\href{http://arxiv.org/abs/0811.1989}{{\tt arXiv:0811.1989}}].

\bibitem{Kaloper:2011jz}
N.~Kaloper, A.~Lawrence, and L.~Sorbo, {\it {An Ignoble Approach to Large Field
  Inflation}},  {\em JCAP} {\bf 1103} (2011) 023,
  [\href{http://arxiv.org/abs/1101.0026}{{\tt arXiv:1101.0026}}].

\bibitem{Yonekura:2014oja}
K.~Yonekura, {\it {Notes on natural inflation}},  {\em JCAP} {\bf 1410} (2014),
  no.~10 054, [\href{http://arxiv.org/abs/1405.0734}{{\tt arXiv:1405.0734}}].
  
\bibitem{Antusch:2017flz}
S.~Antusch, F.~Cefala, S.~Krippendorf, F.~Muia, S.~Orani, and F.~Quevedo, {\it
  {Oscillons from String Moduli}},  {\em JHEP} {\bf 01} (2018) 083,
  [\href{http://arxiv.org/abs/1708.08922}{{\tt arXiv:1708.08922}}].
  
\bibitem{Zhou:2013tsa}
  S.~Y.~Zhou, E.~J.~Copeland, R.~Easther, H.~Finkel, Z.~G.~Mou and P.~M.~Saffin,
  {\it {Gravitational Waves from Oscillon Preheating}},
  {\em JHEP}, {\bf 1310} (2013) 026
  [[\href{http://arxiv.org/abs/1304.6094}{{\tt arXiv:1304.6094}}].
  
\bibitem{Amin:2018xfe}
  M.~A.~Amin, J.~Braden, E.~J.~Copeland, J.~T.~Giblin, C.~Solorio, Z.~J.~Weiner and S.~Y.~Zhou,
  {\it {Gravitational waves from asymmetric oscillon dynamics?}}
  {\em Phys.\ Rev.}, {\bf D98} (2018) 024040
  [\href{http://arxiv.org/abs/1803.08047}{{\tt arXiv:1803.08047}}] 


\bibitem{Kallosh:2013hoa}
R.~Kallosh and A.~Linde, {\it {Universality Class in Conformal Inflation}},
  {\em JCAP} {\bf 1307} (2013) 002, [\href{http://arxiv.org/abs/1306.5220}{{\tt
  arXiv:1306.5220}}].

\bibitem{Kallosh:2013tua}
R.~Kallosh, A.~Linde, and D.~Roest, {\it {Universal Attractor for Inflation at
  Strong Coupling}},  {\em Phys. Rev. Lett.} {\bf 112} (2014), no.~1 011303,
  [\href{http://arxiv.org/abs/1310.3950}{{\tt arXiv:1310.3950}}].

\bibitem{Kallosh:2013yoa}
R.~Kallosh, A.~Linde, and D.~Roest, {\it {Superconformal Inflationary
  $\alpha$-Attractors}},  {\em JHEP} {\bf 11} (2013) 198,
  [\href{http://arxiv.org/abs/1311.0472}{{\tt arXiv:1311.0472}}].

\bibitem{Nomura:2017ehb}
Y.~Nomura, T.~Watari, and M.~Yamazaki, {\it {Pure Natural Inflation}},  {\em
  Phys. Lett.} {\bf B776} (2018) 227--230,
  [\href{http://arxiv.org/abs/1706.08522}{{\tt arXiv:1706.08522}}].

\bibitem{Nomura:2017zqj}
Y.~Nomura and M.~Yamazaki, {\it {Tensor Modes in Pure Natural Inflation}},
  {\em Phys. Lett.} {\bf B780} (2018) 106--110,
  [\href{http://arxiv.org/abs/1711.10490}{{\tt arXiv:1711.10490}}].

\bibitem{Landete:2017amp}
A.~Landete, F.~Marchesano, G.~Shiu, and G.~Zoccarato, {\it {Flux Flattening in
  Axion Monodromy Inflation}},  {\em JHEP} {\bf 06} (2017) 071,
  [\href{http://arxiv.org/abs/1703.09729}{{\tt arXiv:1703.09729}}].
  
\bibitem{Hong:2017ooe}
  J.~P.~Hong, M.~Kawasaki and M.~Yamazaki,
  {\it {Oscillons from Pure Natural Inflation}},
  {\em Phys.\ Rev.} {\bf D98} (2018) no.4,  043531,
 [\href{http://arxiv.org/abs/1711.10496}{{\tt arXiv:1711.10496}}]


\bibitem{Lozanov:2019ylm}
K.~D. Lozanov and M.~A. Amin, {\it {Gravitational perturbations from oscillons
  and transients after inflation}},
  \href{http://arxiv.org/abs/1902.06736}{{\tt arXiv:1902.06736}}.

\bibitem{Fukunaga:2019unq}
H.~Fukunaga, N.~Kitajima, and Y.~Urakawa, {\it {Efficient self-resonance
  instability from axions}},  \href{http://arxiv.org/abs/1903.02119}{{\tt
  arXiv:1903.02119}}.

\bibitem{Cotner:2018vug}
E.~Cotner, A.~Kusenko, and V.~Takhistov, {\it {Primordial Black Holes from
  Inflaton Fragmentation into Oscillons}},  {\em Phys. Rev.} {\bf D98} (2018),
  no.~8 083513, [\href{http://arxiv.org/abs/1801.03321}{{\tt
  arXiv:1801.03321}}].

\bibitem{Masso:2005zg}
E.~Masso, F.~Rota, and G.~Zsembinszki, {\it {Scalar field oscillations
  contributing to dark energy}},  {\em Phys. Rev.} {\bf D72} (2005) 084007,
  [\href{http://arxiv.org/abs/astro-ph/0501381}{{\tt astro-ph/0501381}}].

\bibitem{whittaker1996course}
E.~Whittaker and G.~Watson, {\em A Course of Modern Analysis}.
\newblock Cambridge Mathematical Library. Cambridge University Press, 1996.

\bibitem{Amin:2011hu}
M.~A. Amin, P.~Zukin, and E.~Bertschinger, {\it {Scale-Dependent Growth from a
  Transition in Dark Energy Dynamics}},  {\em Phys. Rev.} {\bf D85} (2012)
  103510, [\href{http://arxiv.org/abs/1108.1793}{{\tt arXiv:1108.1793}}].

\bibitem{Amin:2019ums}
M.~A. Amin and P.~Mocz, {\it {Formation, Gravitational Clustering and
  Interactions of Non-relativistic Solitons in an Expanding Universe}},
  \href{http://arxiv.org/abs/1902.07261}{{\tt arXiv:1902.07261}}.
  
\bibitem{Tanabashi:2018oca}
  M.~Tanabashi {\it et al.} [Particle Data Group],
  Phys.\ Rev.\ D {\bf 98} (2018) no.3,  030001.
  doi:10.1103/PhysRevD.98.030001



\bibitem{Dvali:2017ruz}
G.~Dvali and S.~Zell, {\it {Classicality and Quantum Break-Time for Cosmic
  Axions}},  {\em JCAP} {\bf 1807} (2018), no.~07 064,
  [\href{http://arxiv.org/abs/1710.00835}{{\tt arXiv:1710.00835}}].

\bibitem{Gleiser:2009ys}
M.~Gleiser and D.~Sicilia, {\it {A General Theory of Oscillon Dynamics}},  {\em
  Phys. Rev.} {\bf D80} (2009) 125037,
  [\href{http://arxiv.org/abs/0910.5922}{{\tt arXiv:0910.5922}}].

\bibitem{Hindmarsh:2006ur}
M.~Hindmarsh and P.~Salmi, {\it {Numerical investigations of oscillons in 2
  dimensions}},  {\em Phys. Rev.} {\bf D74} (2006) 105005,
  [\href{http://arxiv.org/abs/hep-th/0606016}{{\tt hep-th/0606016}}].

\bibitem{Liu:2018rrt}
J.~Liu, Z.-K. Guo, R.-G. Cai, and G.~Shiu, {\it {Gravitational wave production
  after inflation with cuspy potentials}},
  \href{http://arxiv.org/abs/1812.09235}{{\tt arXiv:1812.09235}}.

\bibitem{Akrami:2018odb}
{\bf Planck} Collaboration, Y.~Akrami et~al., {\it {Planck 2018 results. X.
  Constraints on inflation}},  \href{http://arxiv.org/abs/1807.06211}{{\tt
  arXiv:1807.06211}}.

\bibitem{Bozek:2014uqa}
B.~Bozek, D.~J.~E. Marsh, J.~Silk, and R.~F.~G. Wyse, {\it {Galaxy
  UV-luminosity function and reionization constraints on axion dark matter}},
  {\em Mon. Not. Roy. Astron. Soc.} {\bf 450} (2015), no.~1 209--222,
  [\href{http://arxiv.org/abs/1409.3544}{{\tt arXiv:1409.3544}}].

\bibitem{Hlozek:2014lca}
R.~Hlozek, D.~Grin, D.~J.~E. Marsh, and P.~G. Ferreira, {\it {A search for
  ultralight axions using precision cosmological data}},  {\em Phys. Rev.} {\bf
  D91} (2015), no.~10 103512, [\href{http://arxiv.org/abs/1410.2896}{{\tt
  arXiv:1410.2896}}].

\bibitem{Ikeda:2017qev}
T.~Ikeda, C.-M. Yoo, and V.~Cardoso, {\it {Self-gravitating oscillons and new
  critical behavior}},  {\em Phys. Rev.} {\bf D96} (2017), no.~6 064047,
  [\href{http://arxiv.org/abs/1708.01344}{{\tt arXiv:1708.01344}}].
  
  
\bibitem{Carr:1997cn}
  B.~J.~Carr and M.~Sakellariadou,
  Astrophys.\ J.\  {\bf 516} (1999) 195.
  doi:10.1086/307071

\bibitem{Carr:2018rid}
B.~Carr and J.~Silk, {\it {Primordial Black Holes as Generators of Cosmic
  Structures}},  {\em Mon. Not. Roy. Astron. Soc.} {\bf 478} (2018), no.~3
  3756--3775, [\href{http://arxiv.org/abs/1801.00672}{{\tt arXiv:1801.00672}}].

\bibitem{Arvanitaki:2010sy}
A.~Arvanitaki and S.~Dubovsky, {\it {Exploring the String Axiverse with
  Precision Black Hole Physics}},  {\em Phys. Rev.} {\bf D83} (2011) 044026,
  [\href{http://arxiv.org/abs/1004.3558}{{\tt arXiv:1004.3558}}].

\bibitem{Arvanitaki:2014wva}
A.~Arvanitaki, M.~Baryakhtar, and X.~Huang, {\it {Discovering the QCD Axion
  with Black Holes and Gravitational Waves}},  {\em Phys. Rev.} {\bf D91}
  (2015), no.~8 084011, [\href{http://arxiv.org/abs/1411.2263}{{\tt
  arXiv:1411.2263}}].

\bibitem{Arvanitaki:2016qwi}
A.~Arvanitaki, M.~Baryakhtar, S.~Dimopoulos, S.~Dubovsky, and R.~Lasenby, {\it
  {Black Hole Mergers and the QCD Axion at Advanced LIGO}},  {\em Phys. Rev.}
  {\bf D95} (2017), no.~4 043001, [\href{http://arxiv.org/abs/1604.03958}{{\tt
  arXiv:1604.03958}}].

\bibitem{Gruzinov:2016hcq}
A.~Gruzinov, {\it {Black Hole Spindown by Light Bosons}},
  \href{http://arxiv.org/abs/1604.06422}{{\tt arXiv:1604.06422}}.

\bibitem{Helfer:2016ljl}
T.~Helfer, D.~J.~E. Marsh, K.~Clough, M.~Fairbairn, E.~A. Lim, and R.~Becerril,
  {\it {Black hole formation from axion stars}},  {\em JCAP} {\bf 1703} (2017),
  no.~03 055, [\href{http://arxiv.org/abs/1609.04724}{{\tt arXiv:1609.04724}}].

\bibitem{Clough:2018exo}
K.~Clough, T.~Dietrich, and J.~C. Niemeyer, {\it {Axion star collisions with
  black holes and neutron stars in full 3D numerical relativity}},  {\em Phys.
  Rev.} {\bf D98} (2018), no.~8 083020,
  [\href{http://arxiv.org/abs/1808.04668}{{\tt arXiv:1808.04668}}].

\bibitem{Clough:2019jpm}
K.~Clough, P.~G. Ferreira, and M.~Lagos, {\it {On the growth of massive scalar
  hair around a Schwarzschild black hole}},
  \href{http://arxiv.org/abs/1904.12783}{{\tt arXiv:1904.12783}}.

\bibitem{inpreparation}
J.~Olle, O.~Pujolas, and F.~Rompineve, {\it {to appear}}, .

\bibitem{Khmelnitsky:2013lxt}
A.~Khmelnitsky and V.~Rubakov, {\it {Pulsar timing signal from ultralight
  scalar dark matter}},  {\em JCAP} {\bf 1402} (2014) 019,
  [\href{http://arxiv.org/abs/1309.5888}{{\tt arXiv:1309.5888}}].

\bibitem{DeMartino:2017qsa}
I.~De~Martino, T.~Broadhurst, S.~H. Henry~Tye, T.~Chiueh, H.-Y. Schive, and
  R.~Lazkoz, {\it {Recognizing Axionic Dark Matter by Compton and de Broglie
  Scale Modulation of Pulsar Timing}},  {\em Phys. Rev. Lett.} {\bf 119}
  (2017), no.~22 221103, [\href{http://arxiv.org/abs/1705.04367}{{\tt
  arXiv:1705.04367}}].

\end{thebibliography}
\end{document}